\documentclass[aps,pre,onecolumn,notitlepage,11pt,tightenlines,superscriptaddress,showkeys,longbibliography,nofootinbib,floatfix]{revtex4-2}

\usepackage{amsmath}
\usepackage{bbold}
\usepackage{graphicx}
\usepackage[colorlinks=true,allcolors=blue]{hyperref}

\DeclareMathOperator*{\argmax}{argmax}

\makeatletter
\def\frontmatter@preabstractspace{24pt}
\def\@keys@name{\textbf{Keywords:} }
\makeatother

\begin{document}

\title{Poisson Log-Normal Process for Count Data Prediction}

\author{Anushka Saha}
\affiliation{Department of Physics \& Astronomy, Rutgers, The State University of New Jersey,
136 Frelinghuysen Rd., Piscataway, NJ 08854, U.S.A.}

\author{Abhijith Gandrakota}
\affiliation{Fermi National Accelerator Laboratory, Batavia, IL 60510, U.S.A.}

\author{Alexandre V. Morozov}
\email[Author to whom any correspondence should be addressed. E-mail: ]{morozov@physics.rutgers.edu}
\affiliation{Department of Physics \& Astronomy, Rutgers, The State University of New Jersey,
136 Frelinghuysen Rd., Piscataway, NJ 08854, U.S.A.}

\begin{abstract}
\noindent
Modeling count data is important in physics and other scientific disciplines, where measurements often involve discrete, non-negative quantities such as photon or neutrino detection events. Traditional parametric approaches can be trained to generate integer-count predictions but may struggle with capturing complex, non-linear dependencies often observed in the data. Gaussian process (GP) regression provides a robust non-parametric alternative to modeling continuous data; however, it cannot generate integer outputs. Here, we
model the Poisson rate as the output of a latent GP,
deriving a closed-form Poisson Log-Normal (PoLoN) predictive distribution by approximating the posterior over the latent log-rate and propagating its uncertainty through the Poisson likelihood. We then extend PoLoN to a signal-background model, PoLoN-SB, which separates a smoothly varying GP background from an explicitly parameterized localized signal and estimates signal strength, location, and width. Since the signal location is often unknown in advance, we introduce a search procedure that scans candidate regions across the input domain, locating and characterizing the signal automatically. We evaluate our approach on synthetic datasets with known signal properties and on real-world data, including the open CERN dataset which was used to detect the Higgs boson at the Large Hadron Collider. Our results show that PoLoN and PoLoN-SB offer a practical non-parametric framework for predicting, analyzing, and extracting localized signals from integer-valued count data.
\end{abstract}


\maketitle

\section{Introduction}

In a wide range of scientific disciplines, including experimental particle physics, materials science, chemistry, and astrophysics, measurements naturally take the form of non-negative integer counts. These counts are often reported as a function of variables such as energy, wavelength, frequency, or time. For example, in the Higgs boson search at the Large Hadron Collider (LHC), the data represent the number of di-photon events per energy bin~\cite{ATLAS2012Higgs,CMS2012Higgs}, where the energy corresponds to the invariant mass of the di-photon system, $m_{\gamma\gamma}$. In electron energy loss spectroscopy (EELS), measurements consist of the number of electrons detected per energy-loss bin~\cite{Egerton2005EELSReview,Colliex2022EELSReview}, while in X-ray photoelectron spectroscopy (XPS), photoelectron counts are measured as a function of binding energy~\cite{Isaacs2021XPSReview,Krishna2022XPSPedagogy,Bagus2024XPSReview}. Similar count-based measurements arise in neutrino detection~\cite{Abbasi2010TimeIntegratedSources,Abbasi2011AtmosphericEnergySpectrum,IceCube2013Science,Aartsen2014HighEnergyNeutrinos} and astrophysical observations, including exoplanet searches~\cite{Borucki2010Kepler,Ricker2015TESS} as well as optical/UV~\cite{GalYam2016SN,Ivezic2019LSST} and high-energy transient detection~\cite{Meegan2009FermiGBM,Evans2009SwiftXRT}.

Across these applications, a central data analysis problem is to distinguish localized signals from a smooth, slowly varying background, often in the presence of substantial statistical noise. A useful model should therefore preserve the discrete nature of the observations, provide a flexible representation of the underlying background, and quantify predictive uncertainty. The latter is particularly important when determining whether a localized deviation is statistically significant. Related applications include de-noising, in which a smoothly varying prediction is constructed from a limited number of noisy observations~\cite{Starck1998Multiscale,Frate:2017mai}.

Numerous computational approaches have been developed for signal-background decomposition and background removal, including asymmetric least squares (ALS) and its variants~\cite{Baek2015arPLS}, Poisson wavelet methods~\cite{Starck1998Multiscale}, and explicit Bernstein or polynomial background models~\cite{CMS2012Higgs}. Traditional statistical approaches to count data include Poisson and negative binomial regression~\cite{NelderWedderburn1972,CameronTrivedi1998,Hutchinson2005,White1996}. While these techniques are useful in many settings, parametric background models require a specified functional form or set of features and can become restrictive when the background has complex or spatially varying structure~\cite{Mehta:2018dln,Rocks2022}. Moreover, many commonly used approaches for background removal do not explicitly incorporate the integer-valued nature of the observations, which can become important when event counts are small.

Gaussian processes (GPs) provide a flexible non-parametric framework for modeling smooth functions while quantifying predictive uncertainty~\cite{Rasmussen2005,Bishop2006,cunningham-2012}. Their use has been explored in astrophysics~\cite{ForemanMackey2017,Iyer_2019,Gordon2020,Suzanne2023}, gravitational-wave detection~\cite{Moore:2015sza}, and high-energy particle physics~\cite{Frate:2017mai,gandrakota2023}. In particular, GP-based approaches have been used to model smooth backgrounds in LHC data and identify statistically significant deviations from those backgrounds~\cite{gandrakota2023}. Kernel construction and learning can further increase the flexibility of GP models, with composite-kernel approaches developed for Support Vector Machines~\cite{inproceeding}, Relevance Vector Machines~\cite{RVM}, and GP regression~\cite{pmlr-v28-duvenaud13,pmlr-v28-wilson13}.

For count data, a GP can be used as a latent process governing the intensity of a Poisson probability distribution. This construction results in Cox processes and, in particular, the log-Gaussian Cox process (LGCP), in which the logarithm of the Poisson intensity is modeled as a GP~\cite{moller1998log,Frid_2026}. The exponential transformation of the latent process ensures a positive intensity, while the Poisson likelihood preserves the integer-valued nature of the observations.
In this work, we develop a specific predictive formulation of this established Poisson-GP framework. We model the logarithm of the Poisson rate as a GP and explicitly derive the corresponding Poisson Log-Normal (PLN) predictive distribution~\cite{izsak2008maximum,shaban2018poisson} by integrating over the posterior uncertainty in the latent GP.

Because the Poisson likelihood makes this posterior non-Gaussian, approximate inference is required~\cite{Bayesian_inference_latent_Gaussian_models}; Monte Carlo methods, as used in a recent LGCP application to collider data~\cite{Frid_2026}, and variational methods~\cite{pmlr-v37-lloyd15} are potential alternatives.
Here, we use a Laplace (saddle-point) approximation, which is efficient enough
for the inference to be repeated over many signal configurations and window
positions. The posterior uncertainty is propagated analytically through the exponential link and the Poisson likelihood, resulting in an explicit formula
for the predictive distribution and its moments.
We refer to our formulation as the Poisson Log-Normal process (PoLoN).

The PoLoN formulation provides a flexible probabilistic description of count data and forms the basis for the signal-background decomposition model developed in this work. Specifically, we represent the total Poisson rate as the combination of a smooth background described by the latent GP and an explicitly parameterized localized signal. We refer to this construction as Poisson Log-Normal Signal-Background (PoLoN-SB). For a one-dimensional (1D) input variable, the signal is represented by a Gaussian function characterized by its amplitude, location, and width. The signal is introduced directly at the level of the Poisson rate, rather than as a mean function of the latent GP, allowing the localized component to be explicitly parameterized while the background retains the flexible GP representation. The resulting model permits simultaneous estimation of the signal parameters and the complex background.
A further challenge in signal-search problems is that the signal location is generally not known \emph{a priori}.
We therefore develop a systematic signal-search algorithm based on a sliding window.
This procedure simultaneously addresses signal detection and parameter estimation without requiring the signal position to be known in advance.

We evaluate the method using controlled synthetic datasets for which the signal and background properties are known. We vary the signal strength, signal location, and number of observations available at each input location, quantifying the recovery of the signal amplitude, location, and width.
We demonstrate our method on real-world count data, including the open CERN dataset associated with the Higgs boson search at the LHC~\cite{ATL-OREACH-PUB-2020-001,CMS2012Higgs}. In summary,
the main contribution of our work is a signal-search framework built on the established Poisson-GP probabilistic construction. Unlike previous approaches, our method utilizes an explicit PLN predictive distribution.

\section{Materials and Methods}

\subsection{Predictive distribution}
\label{sec:methode_1}

Let us consider a training dataset
$\mathcal{D} = \{(\vec{X}_n,\,\{t_{n,i}\}_{i=1}^{N_n})\}_{n=1}^{M}$,
where $\vec{X}_n$ are input feature vectors and $\{t_{n,i}\}_{i=1}^{N_n}$
are $N_n$ independent Poisson-distributed observations collected at
location $\vec{X}_n$. We collect the observation counts per location into
the vector $\vec{N} = [N_1\;\ldots\;N_M]^T$, where $N_n \geq 1$ is allowed
to vary across locations. The total number of observations is
$N = \sum_{n=1}^{M}N_n = \vec{1}^T\vec{N}$.

We assume that the observed counts are Poisson-distributed with rate
parameters $\{\alpha(\vec{X}_n)\}$. To ensure the positivity of Poisson
rates, we express them as $\alpha(\vec{X}_n) = e^{\lambda(\vec{X}_n)}$,
where $\lambda(\vec{X}_n)$ is the latent log-rate at location $\vec{X}_n$.
Because all $N_n$ replicates at location $\vec{X}_n$ arise from repeated
measurements of the same physical process, they share the same latent
log-rate $\lambda_n \equiv \lambda(\vec{X}_n)$. The probability of a
single observation $t_{n,i}$ given this shared log-rate is therefore
\begin{equation} \label{Pois:exp}
    p(t_{n,i}|\lambda_n)
    = \frac{e^{t_{n,i}\lambda_n}e^{-e^{\lambda_n}}}{t_{n,i}!},
    \quad t_{n,i}=0,1,2,\dots
\end{equation}

We collect the $M$ latent log-rates into the vector
$\vec{\lambda}_M = [\lambda(\vec{X}_1)\;\ldots\;\lambda(\vec{X}_M)]^T$.
Let $t_{M+1}$ denote a new target variable corresponding to an input
feature vector $\vec{X}_{M+1}$ not included in the training set; we assume
it is also Poisson-distributed with log-rate $\lambda(\vec{X}_{M+1})$.
Our goal is to determine the predictive distribution
$p(t_{M+1}|\vec{t}_N)$, where
$\vec{t}_N = [t_{1,1}\;\ldots\;t_{M,N_M}]^T$
is the vector of all training observations (we omit the conditional
dependence on the input locations for brevity).

We define a vector
$\vec{\lambda}_{M+1} =
[\lambda(\vec{X}_1)\;\ldots\;\lambda(\vec{X}_{M+1})]^T$
which comprises Poisson log-rates for all $M+1$ locations.
We model the Poisson log-rate as a Gaussian process, following the
standard construction in which a Gaussian latent process governs the
intensity of a Poisson model. The joint probability of
$\vec{\lambda}_{M+1}$ is therefore given by the multivariate Gaussian
distribution:
\begin{equation} \label{Lograte:joint}
    p(\vec{\lambda}_{M+1}) = \mathcal{N}(\vec{\lambda}_{M+1}|\vec{0},C_{M+1}),
\end{equation}
where $\vec{0}$ is a null vector of size $M+1$ and the $(n,m)$ element of
the $(M+1)\times(M+1)$ covariance matrix $C_{M+1}$ is given by
\begin{equation} \label{Lograte:cov}
    C_{M+1,nm} = k_{\vec{\theta}}(\vec{X}_n,\vec{X}_m) + \epsilon\,\delta_{nm},
\end{equation}
where $k_{\vec{\theta}}(\vec{X},\vec{Y})$ is a pre-specified kernel function
with hyperparameters $\vec{\theta}$, $\epsilon>0$ is a small positive constant
added for regularization, and $\delta_{nm}$ is the Kronecker symbol.

Kernel functions are used to quantify the similarity or the amount of
covariance between two input feature vectors; several standard families of
kernel functions and their combinations are available in the GP and machine
learning literature. Here, we employ two kernel function types: the Radial
Basis Function (RBF) kernel and the linear kernel. The RBF kernel is defined
as
\begin{equation} \label{RBF:kernel}
    k_{\vec{\theta}}(\vec{X},\vec{Y}) =
    \gamma\exp\!\left(-\frac{\|\vec{X}-\vec{Y}\|^2}{2\sigma_l^2}\right),
\end{equation}
where $\vec{\theta}=(\sigma_l,\gamma)$, with $\sigma_l$ denoting the length scale and
$\gamma$ the amplitude of the kernel covariance function. The linear
kernel is formulated as
\begin{equation} \label{lin:kernel}
    k_{\vec{\theta}}(\vec{X},\vec{Y}) = \theta_1(\vec{X}\cdot\vec{Y})+\theta_2,
\end{equation}
where $\vec{\theta}=(\theta_1,\theta_2)$, with $\theta_1$ denoting the scaling
term and $\theta_2$ the offset term.

Next, we write the predictive distribution as
\begin{equation} \label{Pred:1}
    p(t_{M+1}|\vec{t}_N)
    = \int d\lambda_{M+1}\;p(t_{M+1}|\lambda_{M+1})\,p(\lambda_{M+1}|\vec{t}_N),
\end{equation}
where $p(t_{M+1}|\lambda_{M+1})$ is given by Eq.~\eqref{Pois:exp} with
$n\to M+1$, and
\begin{equation} \label{P:lambda:t}
    p(\lambda_{M+1}|\vec{t}_N)
    = \int d\vec{\lambda}_M\;p(\lambda_{M+1}|\vec{\lambda}_M)\,
      p(\vec{\lambda}_M|\vec{t}_N).
\end{equation}

The first probability under the integral in the right-hand side of
Eq.~\eqref{P:lambda:t} is a conditional Gaussian distribution obtained from
Eq.~\eqref{Lograte:joint}~\cite{Bishop2006}:
\begin{equation} \label{P:lambda:lambda}
    p(\lambda_{M+1}|\vec{\lambda}_M)
    = \mathcal{N}\!\left(\lambda_{M+1}\,\Big|\,
      \vec{k}^T C_M^{-1}\vec{\lambda}_M,\;
      c - \vec{k}^T C_M^{-1}\vec{k}\right),
\end{equation}
where $C_M$ is the $M\times M$ covariance matrix defined as in
Eq.~\eqref{Lograte:cov} for the $M$ unique training locations,
\begin{equation} \label{little:c}
    c = k_{\vec{\theta}}(\vec{X}_{M+1},\vec{X}_{M+1})+\epsilon
\end{equation}
is the value of the kernel function for the new datapoint, and
\begin{equation} \label{k:vec}
    \vec{k} = \bigl[k_{\vec{\theta}}(\vec{X}_1,\vec{X}_{M+1})\;\ldots\;
    k_{\vec{\theta}}(\vec{X}_M,\vec{X}_{M+1})\bigr]^T
\end{equation}
is a vector of kernel function values for the new datapoint vs.\ $M$ training
locations.

Next, we focus on computing the second probability under the integral in the
right-hand side of Eq.~\eqref{P:lambda:t}, $p(\vec{\lambda}_M|\vec{t}_N)$.
The posterior $p(\vec{\lambda}_M|\vec{t}_N)$ is not Gaussian because of the
Poisson likelihood and is therefore approximated using a Laplace (saddle-point)
approximation. By Bayes' theorem,
\begin{equation} \label{lambda:t:N}
    p(\vec{\lambda}_M|\vec{t}_N)
    = \frac{p(\vec{t}_N|\vec{\lambda}_M)\,p(\vec{\lambda}_M)}{p(\vec{t}_N)}.
\end{equation}
Here, the joint probability
\begin{equation} \label{lambda:t:N:1}
    p(\vec{\lambda}_M) = \mathcal{N}(\vec{\lambda}_M|\vec{0},C_M)
\end{equation}
is generated by the Gaussian process (cf.\ Eq.~\eqref{Lograte:joint}) and
the joint likelihood is the product over all $M$ locations:
\begin{equation} \label{lambda:t:N:2}
    p(\vec{t}_N|\vec{\lambda}_M)
    = \prod_{n=1}^{M}\prod_{i=1}^{N_n}
      \frac{e^{t_{n,i}\lambda_n}\,e^{-e^{\lambda_n}}}{t_{n,i}!}.
\end{equation}
Because $\lambda_n = \lambda(\vec{X}_n)$ is the same for all $N_n$ replicates at location
$\vec{X}_n$, the inner product over $i$ simplifies:
\begin{equation} \label{inner:simplified}
    \prod_{i=1}^{N_n}
    \frac{e^{t_{n,i}\lambda_n}\,e^{-e^{\lambda_n}}}{t_{n,i}!}
    =
    \frac{e^{T_n\lambda_n}\,e^{-N_n e^{\lambda_n}}}
         {\prod_{i=1}^{N_n}t_{n,i}!},
\end{equation}
where
\begin{equation} \label{Tn:def}
    T_n = \sum_{i=1}^{N_n} t_{n,i}
\end{equation}
is the total count at location $\vec{X}_n$. Substituting Eq.~\eqref{inner:simplified} into Eq.~\eqref{lambda:t:N:2}, we find
\begin{equation} \label{joint:lik}
    p(\vec{t}_N|\vec{\lambda}_M)
    = \prod_{n=1}^{M}
      \frac{e^{T_n\lambda_n}\,e^{-N_n e^{\lambda_n}}}
           {\prod_{i=1}^{N_n}t_{n,i}!}.
\end{equation}
Next, we observe that
\begin{equation} \label{log:lambda:t:N}
    \log p(\vec{\lambda}_M|\vec{t}_N)
    = \Psi(\vec{\lambda}_M)
      - \frac{M}{2}\log(2\pi)
      - \frac{1}{2}\log|C_M|
      - \sum_{n=1}^{M}\sum_{i=1}^{N_n}\log(t_{n,i}!)
      - \log p(\vec{t}_N),
\end{equation}
where $|C_M|$ is the determinant of $C_M$ and $\Psi(\vec{\lambda}_M)$ is the
sum of the two terms that depend on the Poisson log-rates:
\begin{equation} \label{Psi:lambda}
    \Psi(\vec{\lambda}_M)
    = -\frac{1}{2}\vec{\lambda}_M^T C_M^{-1}\vec{\lambda}_M
      + \sum_{n=1}^{M}\bigl(T_n\lambda_n - N_n e^{\lambda_n}\bigr).
\end{equation}

To find the mean of the multivariate Gaussian in the Laplace approximation
for $p(\vec{\lambda}_M|\vec{t}_N)$, we need to solve the system of non-linear
equations given by
\begin{equation} \label{Laplace:mean:1}
    \nabla_{\vec{\lambda}_M}\log p(\vec{\lambda}_M|\vec{t}_N)
    \Big|_{\vec{\lambda}_M=\vec{\lambda}_M^\star}
    = \nabla_{\vec{\lambda}_M}\Psi(\vec{\lambda}_M)
    \Big|_{\vec{\lambda}_M=\vec{\lambda}_M^\star}
    = \vec{0},
\end{equation}
which yields
\begin{equation} \label{Laplace:mean:2}
    C_M^{-1}\vec{\lambda}_M^\star = \vec{D}(\vec{\lambda}_M^\star),
\end{equation}
with $\vec{D}(\vec{\lambda}_M) =
[D_1(\lambda_1)\;\ldots\;D_M(\lambda_M)]^T$ and
\begin{equation} \label{residual}
    D_n(\lambda_n) = T_n - N_n e^{\lambda_n},
    \quad n=1,\ldots,M.
\end{equation}
The residual $D_n$ is the difference between the total observed count $T_n$
and the total expected count $N_n e^{\lambda_n}$ at location $\vec{X}_n$.
To find the solution of Eq.~\eqref{Laplace:mean:2},
$\vec{\lambda}_M^\star = C_M\vec{D}(\vec{\lambda}_M^\star)$,
we use the Newton-Raphson (NR) method with an iterative update
defined as
\begin{equation} \label{NR:Laplace}
    \vec{\lambda}_M^{new}
    = \vec{\lambda}_M^{old}
      + H^{-1}(\vec{\lambda}_M^{old})
        \bigl[\vec{D}(\vec{\lambda}_M^{old})
        - C_M^{-1}\vec{\lambda}_M^{old}\bigr].
\end{equation}
Here, $H(\vec{\lambda}_M)$ is an $M\times M$ Hessian matrix given by
\begin{equation} \label{Hess:Laplace}
    H(\vec{\lambda}_M)
    = -\nabla_{\vec{\lambda}_M}\nabla_{\vec{\lambda}_M}
      \log p(\vec{\lambda}_M|\vec{t}_N)
    = -\nabla_{\vec{\lambda}_M}\nabla_{\vec{\lambda}_M}\Psi(\vec{\lambda}_M)
    = C_M^{-1} + W_M(\vec{\lambda}_M),
\end{equation}
where $W_{M,ij}\equiv -\partial D_i/\partial\lambda_j
= \delta_{ij}N_j e^{\lambda_j}$, such that
\begin{equation} \label{W:matrix}
    W_M(\vec{\lambda}_M)
    = \mathrm{diag}(N_1 e^{\lambda_1},\ldots,N_M e^{\lambda_M}).
\end{equation}
The diagonal entry $W_{M,nn} = N_n e^{\lambda_n}$ reflects the curvature of
the Poisson log-likelihood at location $\vec{X}_n$: locations with more
observations contribute a larger entry and yield a sharper posterior for
$\lambda_n$.
Since $W_M$ is positive definite, so is $H$, and the NR
iterations converge to $\vec{\lambda}_M^\star$; the solver settings are given in
the software repository.
The Laplace approximation then yields
\begin{equation} \label{lambda:t:N:Lapl}
    p(\vec{\lambda}_M|\vec{t}_N)
    \simeq \mathcal{N}(\vec{\lambda}_M|\vec{\lambda}_M^\star,
    H^{-1}(\vec{\lambda}_M^\star)).
\end{equation}

Under this Gaussian approximation, the integral in Eq.~\eqref{P:lambda:t}
can be evaluated analytically, yielding
\begin{equation} \label{lambda:tN:final}
    p(\lambda_{M+1}|\vec{t}_N) = \mathcal{N}(\lambda_{M+1}|\mu,\sigma^2),
\end{equation}
where
\begin{eqnarray} \label{mu:sigma}
    \mu &=& \vec{k}^T C_M^{-1}\vec{\lambda}_M^\star
            = \vec{k}^T\vec{D}(\vec{\lambda}_M^\star), \\
    \sigma^2 &=& c - \vec{k}^T C_M^{-1}\vec{k}
                 + \vec{k}^T C_M^{-1}H^{-1}C_M^{-1}\vec{k}
               = c - \vec{k}^T(C_M + W_M^{-1})^{-1}\vec{k}. \nonumber
\end{eqnarray}
Here, $W_M = \mathrm{diag}(N_1 e^{\lambda_1^\star},\ldots,
N_M e^{\lambda_M^\star})$. Note that all matrix operations in
Eqs.~\eqref{mu:sigma} are $M$-dimensional, in contrast to the $N$-dimensional
system that would arise from treating each of the $N$ observations separately.

Eqs.~\eqref{lambda:tN:final} and \eqref{mu:sigma} apply when
$\vec{X}_{M+1} \notin \{\vec{X}_1,\ldots,\vec{X}_M\}$, i.e.\ when the new datapoint
was not seen during training. In this case, $\lambda_{M+1}$ must be treated as a new latent variable.
When $\vec{X}_{M+1} = \vec{X}_m$ for some $m \in \{1,\ldots,M\}$, the
log-rate at the query location is $\lambda_m$, one of the
$M$ components of $\vec{\lambda}_M$.
In this case, $\mu$ and $\sigma^2$ are given by the marginal of the Laplace posterior
(Eq.~\eqref{lambda:t:N:Lapl}) at coordinate $m$:
\begin{equation} \label{mu:sigma:B}
    \mu = \lambda_m^\star,
    \qquad
    \sigma^2 = \bigl[H^{-1}(\vec{\lambda}_M^\star)\bigr]_{mm}.
\end{equation}
The variance is strictly positive for any finite
$N_m$ because $H(\vec{\lambda}_M^\star)$ is positive definite. It decreases
as $N_m$ increases: the diagonal entry $H_{mm} = [C_M^{-1}]_{mm} +
N_m e^{\lambda_m^\star}$ grows with $N_m$, so that $[H^{-1}]_{mm}$ shrinks.
Note that
predicting at a location
already in the training set carries no GP interpolation uncertainty, only the
posterior uncertainty in $\lambda_m$ itself.

Because the approximate posterior for the latent log-rate is Gaussian, the
corresponding Poisson rate $\alpha_{M+1}=e^{\lambda_{M+1}}$ follows a
log-normal distribution:
\begin{equation} \label{alpha:tN:final}
    p(\alpha_{M+1}|\vec{t}_N) =
    \mathrm{LogNormal}(\alpha_{M+1}|\mu,\sigma^2).
\end{equation}
Substituting Eq.~\eqref{alpha:tN:final} into
Eq.~\eqref{Pred:1} gives the Poisson Log-Normal (PLN) predictive distribution:
\begin{equation} \label{Pred:f1}
    p(t_{M+1}|\vec{t}_N)
    = \int_{-\infty}^{\infty}d\lambda_{M+1}\;
      \frac{e^{t_{M+1}\lambda_{M+1}}e^{-e^{\lambda_{M+1}}}}{t_{M+1}!}\;
      \mathcal{N}(\lambda_{M+1}|\mu,\sigma^2)
\end{equation}
or, equivalently,
\begin{equation} \label{Pred:f2}
    p(t_{M+1}|\vec{t}_N)
    = \int_0^\infty d\alpha\;
      \frac{\alpha^{t_{M+1}}e^{-\alpha}}{t_{M+1}!}\;
      \frac{1}{\alpha\sqrt{2\pi\sigma^2}}
      \exp\!\left(-\frac{(\ln\alpha-\mu)^2}{2\sigma^2}\right) \equiv f(t_{M+1};\mu,\sigma^2),
\end{equation}
a convolution between Poisson and log-normal distributions.
The PLN distribution is a discrete probability distribution normalized over
non-negative counts:
\begin{equation} \label{Pred:norm}
    \sum_{t_{M+1}=0}^\infty f(t_{M+1};\mu,\sigma^2) = 1.
\end{equation}
Its first two moments are given by
$\mathcal{M}(\vec{X})\equiv\mathrm{E}[t_{M+1}] = e^{\mu+\sigma^2/2}$
and $V(\vec{X})\equiv\mathrm{Var}[t_{M+1}]
= e^{\mu+\sigma^2/2}+(e^{\sigma^2}-1)e^{2\mu+\sigma^2}$, respectively.
Note that $\mu=\mu(\vec{X})$ and $\sigma=\sigma(\vec{X})$ depend on the
input feature vector $\vec{X}$ for which the prediction is made.

\subsection{Signal-background decomposition}
\label{sec:methode_2}

We extend the predictive formulation to a signal-background model by representing the total Poisson rate as the sum of a GP-parameterized background rate and an explicitly parameterized localized signal. Rather than modifying the mean function of the latent GP, we introduce the signal directly at the level of the Poisson rate. Thus,
the signal contribution is controlled explicitly by a separate set of parameters.

Specifically, we represent the signal by a non-negative function $g_{\vec{B}}(\vec{X})$,
where $\vec{B}$ is a vector of function parameters.
For example, a common choice for $g_{\vec{B}}(\vec{X})$ when $\vec{X}$ is 1D (i.e., $\vec{X} \to X$) is a Gaussian function characterized by the mean $q$,
the standard deviation $u$, and the amplitude $S$:
\begin{equation} \label{g:Gaus}
g_{\vec{B}} (X) = S \exp \left\{-\frac{(X - q)^2}{2 u^2} \right\}.
\end{equation}
Here, the vector of function parameters is given by \(\vec{B} = \{ S, q, u \}\).

Since all $N_n$ replicates at location $\vec{X}_n$
share the same rate, the total Poisson rate at $\vec{X}_n$ is
\begin{equation} \label{alpha:g}
    \alpha_n^{\mathrm{tot}} = e^{\lambda_n} + g_n
\end{equation}
or, equivalently,
\begin{equation} \label{lambda:g:tot}
    \lambda_n^{\mathrm{tot}}
    = \log (e^{\lambda_n} + g_n),
\end{equation}
where we have defined $g_n = g_{\vec{B}}(\vec{X}_n)$ for brevity.

The probability distribution of $\alpha_{M+1}^{\mathrm{tot}}$ is given by
the three-parameter log-normal distribution:
\begin{eqnarray} \label{Lognormal:shifted}
    p(\alpha_{M+1}^{\mathrm{tot}}|\vec{t}_N)
    &=& \mathrm{LogNormal}(\alpha_{M+1}^{\mathrm{tot}}|\mu,\sigma^2,
        g_{\vec{B}}(\vec{X}_{M+1})) \nonumber\\
    &=& \frac{1}{(\alpha_{M+1}^{\mathrm{tot}}-g_{M+1})\sqrt{2\pi\sigma^2}}
        \exp\!\left\{
        -\frac{(\ln(\alpha_{M+1}^{\mathrm{tot}}-g_{M+1})-\mu)^2}{2\sigma^2}
        \right\},
\end{eqnarray}
where $\mu$ and $\sigma^2$ are given by Eq.~\eqref{mu:sigma} and
$g_{M+1}=g_{\vec{B}}(\vec{X}_{M+1})$. Correspondingly, the predictive
distribution is given by
\begin{equation} \label{Pred:shifted}
    p(t_{M+1}|\vec{t}_N)
    = \int_{g_{M+1}}^\infty d\alpha\;
      \frac{\alpha^{t_{M+1}}e^{-\alpha}}{t_{M+1}!}\;
      \frac{1}{(\alpha-g_{M+1})\sqrt{2\pi\sigma^2}}
      \exp\!\left(-\frac{(\ln(\alpha-g_{M+1})-\mu)^2}{2\sigma^2}\right),
\end{equation}
a generalization of Eq.~\eqref{Pred:f2} to the case of non-zero
prior Poisson rates.

\subsection{Hyperparameter optimization}
In this section, we describe optimization of the GP kernel hyperparameters and, for the signal-background model, the signal parameters.

\noindent
\textbf{Kernel hyperparameters.}
Here, we develop a procedure for finding the optimal values of kernel hyperparameters $\vec{\theta}^\star$. Following a standard approach widely used in GP regression~\cite{Bishop2006}, we obtain $\vec{\theta}^\star$ by maximizing the marginal likelihood of the counts observed in the training dataset (cf. Eq.~\eqref{lambda:t:N}):
\begin{equation} \label{L:marginal}
    p(\vec{t}_N | \vec{\theta}) = \int d \vec{\lambda}_{M}~p(\vec{t}_{N} | \vec{\lambda}_{M}) p(\vec{\lambda}_{M} | \vec{\theta}),
\end{equation}
where the dependence on kernel hyperparameters is made explicit for clarity. Since the multi-dimensional integral on the right-hand side of  Eq.~\eqref{L:marginal} cannot be evaluated exactly, we resort once again to the Laplace approximation:
\begin{equation} \label{L:marginal:Laplace}
    \log p(\vec{t}_N|\vec{\theta}) \approx \log p(\vec{\lambda}_M^\star | \vec{\theta}) + \log p(\vec{t}_N|\vec{\lambda}_M^\star) -
    \frac{1}{2} \log(|H(\vec{\lambda}_M^\star,\vec{\theta})|) + \frac{M}{2} \log(2\pi),
\end{equation}
where, as before, the numerical values of $\vec{\lambda}_M^\star$ are found by solving Eq.~\eqref{Laplace:mean:1} (note that $\Psi (\vec{\lambda}_M)$ contains all the $\vec{\lambda}_M$-dependent terms in $\log [p(\vec{\lambda}_M | \vec{\theta}) p(\vec{t}_N|\vec{\lambda}_M)]$).

The optimized hyperparameters $\vec{\theta}^\star$ are found as
\begin{equation} \label{theta:argmax}
\argmax_{\vec{\theta}} \left\{ \Psi (\vec{\lambda}_M^\star (\vec{\theta})) - \frac{1}{2} \log | \mathbb{1}_M + C_M (\vec{\theta}) W_M(\vec{\lambda}_M^\star (\vec{\theta}))|\right\},
\end{equation}
where $\mathbb{1}_M$ is an $M \times M$ unit matrix.

The maximization
in Eq.~\eqref{theta:argmax} is carried out as follows: for a fixed value of $\vec{\theta}$, the corresponding values of
$\vec{\lambda}_M^\star (\vec{\theta})$ are found via NR iterations of Eq.~\eqref{Laplace:mean:1} (cf. Eq.~\eqref{NR:Laplace}).
These values are then substituted into Eq.~\eqref{theta:argmax}, providing a point estimate of the $\text{argmax}$ argument.
The maximization with respect to $\vec{\theta}$ is done using the L-BFGS-B algorithm (Limited-memory Broyden-Fletcher-Goldfarb-Shanno algorithm with Box constraints)~\cite{Broyden1970,Fletcher1970,Goldfarb1970,Shanno1970} implemented in the SciPy library~\cite{byrd1995limited}.
Note that the box constraints in the SciPy L-BFGS-B implementation allow us to restrict any hyperparameter to a prespecified range, ensuring
that the covariance matrix remains positive semi-definite.
In particular, we enforce \(\theta_1 > 0\) and \(\theta_2 > 0\) hyperparameter ranges in the linear kernel (Eq.~\eqref{lin:kernel}).
The L-BFGS-B maximization yields the optimal vector of hyperparameters $\vec{\theta}^\star$ which is used in subsequent predictions.

\vspace{0.1in}
\noindent
\textbf{Signal parameters.} When the prior Poisson rates include a signal contribution, we consider
\begin{equation} \label{L:marginal:prior}
    p(\vec{t}_N|\vec{\theta},\vec{B})
    = \int d\vec{\lambda}_M\;
      p(\vec{t}_N|\vec{\lambda}_M^{\mathrm{tot}})\,
      p(\vec{\lambda}_M|\vec{\theta}),
\end{equation}
where $\vec{\lambda}_M^{\mathrm{tot}} =
[\lambda_1^{\mathrm{tot}}\;\ldots\;\lambda_M^{\mathrm{tot}}]^T$ (cf. Eq.~\eqref{lambda:g:tot}).
Note that we do not modify the zero mean in the joint probability
$p(\vec{\lambda}_M|\vec{\theta})$ (Eq.~\eqref{Lograte:joint}); only the
conditional probability of the observed counts depends on $\vec{B}$
explicitly: $p(\vec{t}_N|\vec{\lambda}_M)\to
p(\vec{t}_N|\vec{\lambda}_M^{\mathrm{tot}})
= p(\vec{t}_N|\vec{\lambda}_M,\vec{B})$.
We carry out the maximization in two separate stages to ensure robust
signal-background separation.

We focus on prior signal functions localized in the feature space, such as
the 1D Gaussian in Eq.~\eqref{g:Gaus}. We divide the training
data into the `background only' and `signal+background' datasets:
$\vec{t}_N \to \{ \vec{t}_w, \vec{t}_o \}$, where $\vec{t}_w$ is a set of
datapoints inside a pre-specified window where the signal is located
and $\vec{t}_o$ is a set of datapoints outside this window.
We fit the parameters in two steps. First, we find $\vec{\theta}^\star$ using the
background-only dataset, $\vec{t}_o$. Second, we find $\vec{B}^\star$ using the
signal+background dataset, $\vec{t}_w$, and maximizing
$\log p(\vec{t}_w|\vec{\theta}^\star,\vec{B})$ with respect to $\vec{B}$,
while keeping $\vec{\theta}=\vec{\theta}^\star$. The kernel hyperparameters
therefore capture the properties of the background, while the prior function
parameters describe the contribution of the signal.

Under the Laplace approximation, Eq.~\eqref{L:marginal:prior} yields:
\begin{equation} \label{L:marginal:Laplace:prior:new}
    \log p(\vec{t}_w|\vec{\theta}^\star,\vec{B})
    \approx \log p(\vec{\lambda}_M^\star|\vec{\theta}^\star)
    + \log p(\vec{t}_w|\vec{\lambda}_M^{\mathrm{tot},\star})
    - \frac{1}{2}\log|H(\vec{\lambda}_M^\star,\vec{\theta}^\star,\vec{B})|
    + \frac{M}{2}\log(2\pi).
\end{equation}
We find $\vec{\lambda}_M^\star$ by solving
\begin{equation} \label{Laplace:mean:1:new}
    \nabla_{\vec{\lambda}_M}\widetilde\Psi(\vec{\lambda}_M,\vec{B})
    \Big|_{\vec{\lambda}_M=\vec{\lambda}_M^\star} = \vec{0},
\end{equation}
where
\begin{equation} \label{Psi:lambda:bkgr:new}
    \widetilde\Psi(\vec{\lambda}_M,\vec{B})
    = -\frac{1}{2}\vec{\lambda}_M^T C_M^{-1}\vec{\lambda}_M
      + \sum_{n=1}^{M}\bigl(\lambda_n^{\mathrm{tot}}T_n
      - N_n e^{\lambda_n^{\mathrm{tot}}}\bigr).
\end{equation}
Here, $M$ refers to the number of positions inside the signal+background window.
Eq.~\eqref{Laplace:mean:1:new} yields
\begin{equation} \label{Laplace:mean:2:new}
    C_M^{-1}\vec{\lambda}_M^\star
    = \widetilde{\vec{D}}(\vec{\lambda}_M^\star,\vec{B}),
\end{equation}
where
\begin{equation} \label{Dtilde}
    \widetilde{D}_n
    = \frac{T_n}{1+g_n e^{-\lambda_n}} - N_n e^{\lambda_n},
    \quad n=1,\ldots,M.
\end{equation}
As before, Eq.~\eqref{Laplace:mean:2:new} can be solved by the NR method,
which requires the Hessian matrix (also used in
Eq.~\eqref{L:marginal:Laplace:prior:new}):
\begin{equation} \label{Hess:Laplace:new}
    H(\vec{\lambda}_M,\vec{\theta}^\star,\vec{B})
    = -\nabla_{\vec{\lambda}_M}\nabla_{\vec{\lambda}_M}
      \widetilde\Psi(\vec{\lambda}_M,\vec{B})
    = C_M^{-1} + \widetilde{W}_M(\vec{\lambda}_M,\vec{B}),
\end{equation}
where
\begin{equation} \label{W:tilde}
    \widetilde{W}_{M,ij}
    = \delta_{ij}\!\left[
      N_j e^{\lambda_j}
      - \frac{T_j g_j e^{-\lambda_j}}{(1+g_j e^{-\lambda_j})^2}
      \right].
\end{equation}

Finally, the optimized signal parameters $\vec{B}^\star$ are found via
\begin{equation} \label{theta:Bmax}
    \argmax_{\vec{B}}\left\{
    \sum_{n=1}^{M}\bigl[T_n\log(e^{\lambda_n^\star}+g_n)
    - N_n(e^{\lambda_n^\star}+g_n)\bigr]
    - \frac{1}{2}\log\bigl|C_M^{-1}
    + \widetilde{W}_M(\vec{\lambda}_M^\star,\vec{B})\bigr|
    \right\}.
\end{equation}
As before, the maximization is done using the L-BFGS-B algorithm, yielding the optimal
vector of signal function parameters $\vec{B}^\star$. Note that, in general,
$\vec{\lambda}_M^\star=\vec{\lambda}_M^\star(\vec{B})$ and thus
$\vec{\lambda}_M^\star$ needs to be recomputed as $\vec{B}$ is updated, even
if $\vec{\theta}=\vec{\theta}^\star$ in Eq.~\eqref{theta:Bmax}.

\vspace{0.1in}\noindent
\textbf{The weak-signal approximation.}
In many situations, we expect the signal to be relatively weak compared to
the background. If the observed Poisson counts are dominated by the background
contribution, we have $g_n e^{-\lambda_n^\star} = g_n/\alpha_n^\star \ll 1$,
$\forall n$. Since $\widetilde{\vec{D}} \to \vec{D}$ and
$\widetilde{W}_M \to W_M$ in this limit, the optimization procedure becomes
considerably simplified. Indeed, $\vec{\lambda}_M^\star(\vec{\theta}^\star)$
can now be found \emph{once} using Eq.~\eqref{Laplace:mean:2}, since it is
independent of $\vec{B}$. Consequently, Eq.~\eqref{theta:Bmax} simplifies to
\begin{equation} \label{theta:Bmax:weak}
    \argmax_{\vec{B}}\left\{
    \sum_{n=1}^{M}\bigl[T_n\log(e^{\lambda_n^\star}+g_n)
    - N_n g_n\bigr]
    \right\}.
\end{equation}
Setting the $\partial/\partial B_i$ derivatives of the $\mathrm{argmax}$
argument in Eq.~\eqref{theta:Bmax:weak} to $0$, we obtain the following
self-consistency condition:
\begin{equation} \label{self:weak}
    \sum_{n=1}^{M}\frac{T_n}{e^{\lambda_n^\star}+g_n}
    \frac{\partial g_n}{\partial B_i}
    = \sum_{n=1}^{M} N_n\frac{\partial g_n}{\partial B_i},
\end{equation}
where $\vec{B}=\vec{B}^\star$ everywhere, resulting in $g_n \to g_n^\star$.
Note that Eq.~\eqref{self:weak}
is consistent with $T_n = N_n(\alpha_n^\star+g_n^\star)$ ($n = 1 \ldots M$),
such that each total count is modeled as $N_n$ times the sum of the background
and signal contributions, with optimized parameters.

\subsection{Detection of signals with unknown location}
\label{sec:methode_3}
The signal-background decomposition and the hyperparameter optimization procedure described above assume that the approximate location of the signal is known \emph{a priori}. However, in many applications the signal position must be inferred directly from the observed count distribution.
To address this limitation, we extend the PoLoN-SB framework by treating the signal location as an additional quantity to be determined.

Specifically, we define a fixed-width signal window (here, we focus on the 1D input variable, the Gaussian signal given by Eq.~\eqref{g:Gaus}, and the RBF kernel
given by Eq.~\eqref{RBF:kernel}). The window is translated across the histogram of counts one bin at a time, with each window position defining a candidate signal region indexed by $j$.
For each window position $j$, the data inside the window are treated as signal+background, while the remaining observations are treated as background only. The signal and background inference steps are repeated independently for every candidate window position
(cf. Eqs.~\eqref{theta:argmax} and \eqref{theta:Bmax}).

We have found it necessary to regularize the background-only marginal log-likelihood with respect to the
RBF length scale $\sigma_l$ (Eq.~\eqref{RBF:kernel}):
\begin{equation}
\log p_j (\vec{t}_o|\vec{\theta}) \to
\log p_j (\vec{t}_o|\vec{\theta}) + R(\sigma_l),
\end{equation}
where $R(\sigma_l) = \lambda\,\sigma\!\left(a(\sigma_l - \sigma_{l,0})\right)$
is a smooth sigmoid-based regularization term for the RBF length scale.
In this work, we use $\lambda=2000$, $a=10$, and $\sigma_{l,0}=0.1$.
Without the constraint, a sufficiently small length scale allows the GP background to
interpolate through the signal window, absorbing the localized
signal into the background fit. The regularization encourages solutions with larger
length scales, ensuring that the background remains smooth and the localized signal
structure is preserved. The regularization is applied only during the kernel hyperparameter optimization step.

The final estimates of the signal and background parameters are obtained through
likelihood-weighted averaging over all candidate windows:
\begin{equation}
\bar{\vec{\theta}} =
\frac{
\sum_j w_j \vec{\theta}^{\star}_{j}
}{
\sum_j w_j
} \text{ and }
\bar{\vec{B}}
=
\frac{
\sum_j w_j \vec{B}^{\star}_{j}
}{
\sum_j w_j
},
\end{equation}
where
\begin{equation}
w_j = \frac{p_j(\vec{t}_N | \vec{\theta}^{\star}_{j},\vec{B}^{\star}_{j})}{\sum_k p_k(\vec{t}_N | \vec{\theta}^{\star}_{k},\vec{B}^{\star}_{k})}.
\end{equation}
Here, $p_j$ refers to the likelihood of the entire dataset conditioned on the
window position $j$. The resulting signal parameter vector, $\bar{\vec{B}} = \{\bar{S},\bar{q},\bar{u}\}$,
provides estimates of the signal strength, location, and width, while
$\bar{\vec{\theta}}$ describes the optimized background.

\section{Results}

We started by validating PoLoN on synthetic
integer-count datasets with known Poisson rate functions. The
validation included 1D rate functions with linear,
polynomial, oscillatory, and exponentially decaying functional forms, as
well as a two-dimensional (2D) rate function with nonlinear dependence
on the input variables. In all cases, PoLoN recovered the underlying
Poisson rates from noisy observations and accurately
interpolated between the sampled input locations (Appendix~\ref{app:validation}).

\subsection{Robust signal-background separation in synthetic data}
\label{sec:results_signal_background}

Next, we have constructed a collection of synthetic datasets inspired by the discovery of the Higgs boson~\cite{ATLAS2012Higgs,CMS2012Higgs,Bass2021_HiggsOutlook} and the ongoing search for new physics at the Large Hadron Collider (LHC) at CERN~\cite{Sopczak2024_BSMHiggs}.
Our objective is to emulate the Higgs boson signal superimposed on a smoothly varying background. We model the background using an exponentially decaying function and represent the Higgs boson signal as a single Gaussian peak~\cite{gandrakota2023}.
Accordingly, we define the Poisson rate function as
\begin{equation} \label{alpha:Higgs}
\alpha(x) = A e^{-\kappa x} + S \exp\left\{-\frac{(x - q)^2}{2u^2}\right\},
\end{equation}
where \(\log A = 11.7\), \(\kappa = 30.6\), \(q = 0.135\), and \(u = 0.004\). These parameters are chosen to mimic the $H \to \gamma \gamma$ invariant mass plot (the $m_{\gamma \gamma}$ mass distribution) reported for the Higgs boson~\cite{CMS2012Higgs}, with $x$ representing the invariant mass of a photon pair, $m_{\gamma \gamma}$, measured in TeV.

In this section, we assume that the approximate location and width of the signal are known through the position of the signal+background window.
We vary the signal strength parameter \(S\) to assess the accuracy of our methodology in detecting the Gaussian peak;
see representative Poisson rate curves in figure~\ref{fig:alpha_signal}.
To create a training dataset, we select $M=40$ equally spaced datapoints in the range \(x \in [0.10, 0.16]\) and use Eq.~\eqref{alpha:Higgs} to generate $N_p$ Poisson counts at each value of $x$.
We use the approach described in Section~\ref{sec:methode_1} (PoLoN) and follow the procedure used in the preceding synthetic examples: (i) generate training set counts using Eq.~\eqref{alpha:Higgs}, (ii) determine the hyperparameters that maximize the posterior log-likelihood, and (iii) use optimal hyperparameter values for prediction. In this approach, signal is not explicitly separated from the background.
In addition, we employ the method described in Section~\ref{sec:methode_2}
(PoLoN-SB) for the same datasets. In this approach, we explicitly predict the signal parameters (signal strength $S$, mean $q$, and standard deviation $u$)
under the assumption that the signal is represented by a single Gaussian peak. These predictions can be compared with the exact values used in the Poisson rate function (Eq.~\eqref{alpha:Higgs}).

\begin{figure}[ht]
    \centering
    \includegraphics[width=\textwidth]{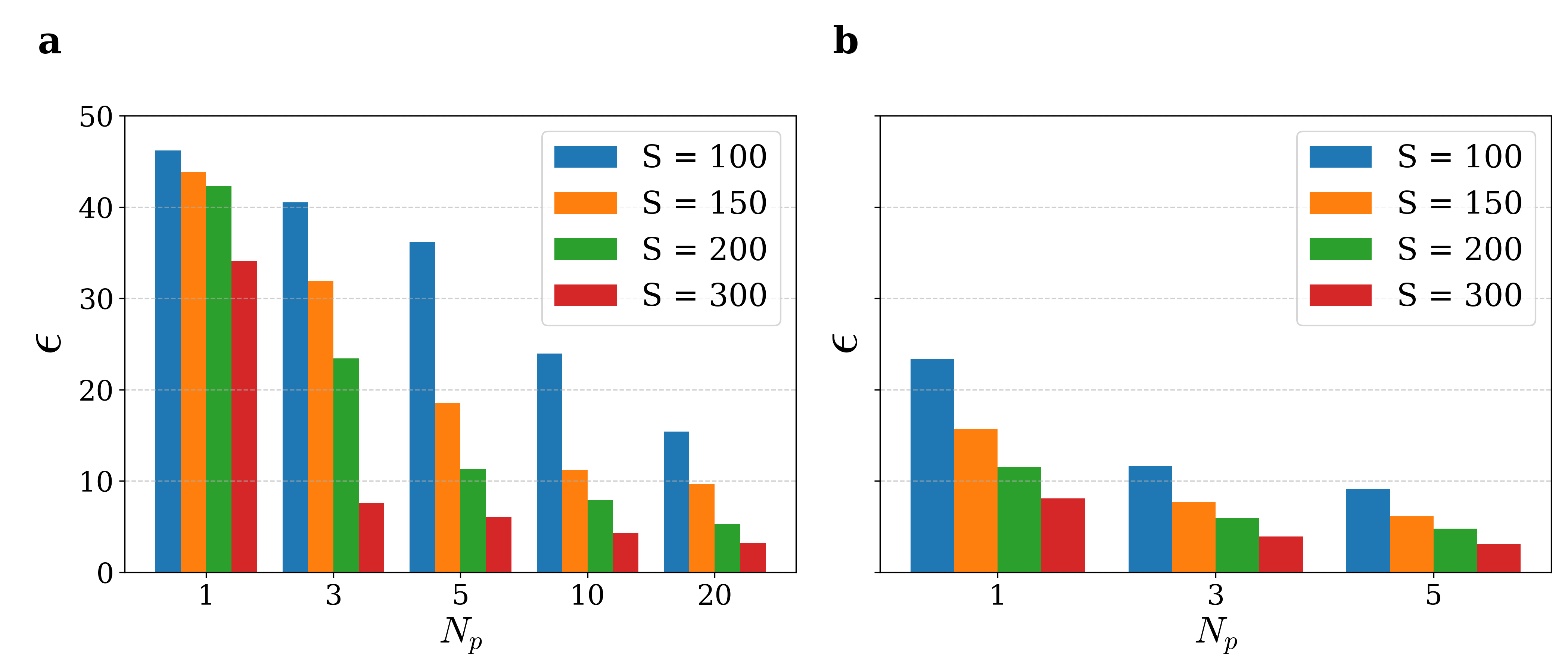}
    \caption{\textbf{Normalized RMSE as a function of $N_p$, the number of datapoints per input location in the training dataset.} Panel (a) shows $\epsilon$ for PoLoN predictions (Section~\ref{sec:methode_1}). Panel (b) shows $\epsilon$ for PoLoN-SB predictions (Section~\ref{sec:methode_2}).
    For each combination of $N_p$ and $S$, the normalized RMSE $\epsilon$ in Eq.~\eqref{rmse:2s} was averaged over $15$ independently generated training datasets.
    }
    \label{fig:rmse_comparison}
\end{figure}

We first assess the ability of both methods to capture the signal by calculating the normalized root-mean-square error (RMSE) in the \(\pm 2u\) window around $q$:
\begin{equation} \label{rmse:2s}
   \epsilon = \frac{\mathrm{RMSE}_{2u}}{S} \times 100,
\end{equation}
where the known signal strength $S$ is used for normalization, such that the RMSE is reported as the percentage of $S$. We use the \(\pm 2u\) window to focus on the accuracy of signal prediction; we already know that exponentially decaying backgrounds can be accurately modeled by our approach (figure~\ref{fig:lin_sin_2}c,d). As before, the RMSE was computed
using the squared differences between the mean of the predictive distribution $\mathcal{M}(x_i)$
and the exact Poisson rates $\alpha(x_i)$ (Eq.~\eqref{alpha:Higgs}), where $x_i$ refer to the $11$ values of $x$ from the training dataset that fall in the \(\pm 2u\) window: $[0.127, 0.143]$, used here as a known signal window.

\begin{figure}[!htb]
    \centering
    \includegraphics[width=\textwidth]{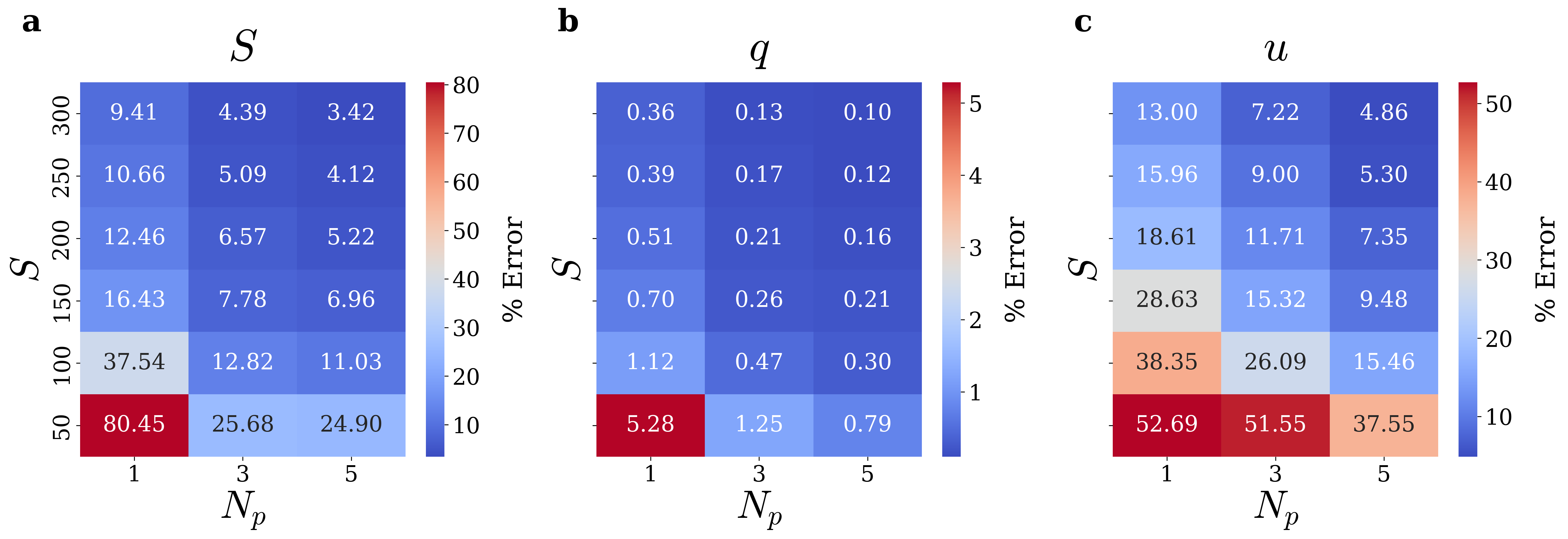}
    \caption{\textbf{Relative percentage errors in Gaussian signal parameter recovery with known signal position.} Panel (a) shows the error in the predicted signal strength $S$, panel (b) shows the error in the predicted mean of the signal $q$, and panel (c) shows the error in the predicted standard deviation of the signal $u$. We used PoLoN-SB (Section~\ref{sec:methode_2}) to predict the signal parameters.
    For each combination of $N_p$ and $S$, the relative percentage error (computed as $\frac{|predicted - true|}{true} \times 100 \%$)
    was averaged over $15$ independently generated training datasets.
    }
    \label{fig:gaussian_parameters}
\end{figure}

Figure~\ref{fig:rmse_comparison} shows the relative error \(\epsilon\) (Eq.~\eqref{rmse:2s}) between the PoLoN and PoLoN-SB methods as a function of the total number of datapoints in the training set. We find that, as expected, the quality of the fit improves with $N_p$ for both methods. Additionally, for a fixed $N_p$, increasing the signal strength enhances the fitting accuracy.
Importantly, for the same values of $N_p$ and $S$
PoLoN-SB demonstrates superior performance compared to PoLoN.

Next, we evaluate how precisely the Gaussian signal parameters are recovered using PoLoN-SB (Section~\ref{sec:methode_2}). Figure~\ref{fig:gaussian_parameters} displays percentage errors in the predicted amplitude $S$, mean $q$, and standard deviation $u$ of the Gaussian peak as a function of the signal strength $S$ and the number of datapoints per input feature $N_p$ in the training set. Thus, we are able to track the accuracy of parameter recovery as $S$ and $N_p$ are varied, complementing the analysis presented in Figure~\ref{fig:rmse_comparison}.
As expected, prediction accuracy improves with increasing both signal strength $S$ and the number of datapoints per input feature $N_p$. The relative errors are comparable in magnitude for $S$ and $u$ (figure~\ref{fig:gaussian_parameters}a,c) and smaller for $q$ (figure~\ref{fig:gaussian_parameters}b). Overall, we conclude that PoLoN-SB can be used to reliably predict signal parameters for relatively weak signal strengths and moderate training set sizes.

\subsection{Search for signals with unknown \emph{a priori} position}
\label{sec:results_unknown_position}

The preceding analysis considered the case in which the approximate signal position and width are known through the use of the $[0.127, 0.143]$ signal window. We now remove the assumption that the window position is specified in advance and consider the more general problem of identifying a localized signal directly from the data. In this setting, PoLoN-SB is applied using the signal-search procedure described in Section~\ref{sec:methode_3}.

\begin{figure}[!htb]
\centering
\includegraphics[width=\textwidth]{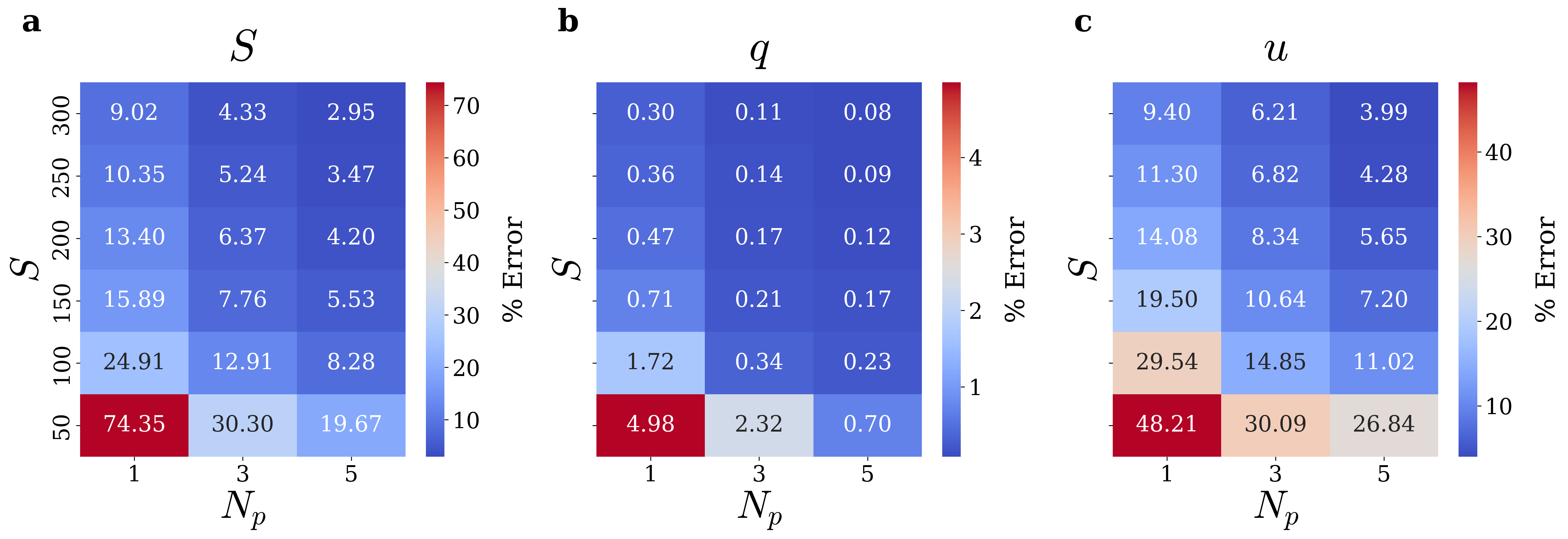}
\caption{\textbf{Relative percentage errors in Gaussian signal parameter recovery with unknown signal position.}
Panel (a) shows the error in the predicted signal strength $S$, panel (b) shows the error in the predicted mean of the signal $q$, and panel (c) shows the error in the predicted standard deviation of the signal $u$,
for a Gaussian signal injected at \(q=0.135\).
We used a PoLoN-SB scan over window positions (Section~\ref{sec:methode_3}) to predict the signal parameters. For each combination of $N_p$ and $S$, the relative percentage error (computed as $\frac{|predicted - true|}{true} \times 100 \%$)
was averaged over $20$ independently generated training datasets.
}
\label{fig:unknown_position_q0135}
\end{figure}

We first consider a Gaussian signal injected at \(q=0.135\).
As before, the analysis is repeated for different values of the signal strength \(S\) and the number of observations per input value \(N_p\). Figure~\ref{fig:unknown_position_q0135} shows the resulting relative percentage errors in the recovered signal parameters. Note that in this case, the recovery of \(q\)
is unaided by the knowledge of the window position -- we have scanned over all sliding
windows of ten-datapoint width, with a one-datapoint step (recall that there are
$M=40$ equally spaced datapoints in the training dataset).
The results in figure~\ref{fig:unknown_position_q0135} indicate that the signal parameters can be recovered reliably without any prior knowledge about the signal width and location. As in the known-window case, the parameter-recovery accuracy improves with increasing \(S\) and \(N_p\).

To test whether the quality of predictions depends on the signal location, we repeat the same analysis for additional values of \(q\). The complete set of parameter-recovery results for four signal locations $q = (0.115,~0.125,~0.135,~0.145)$
is shown in figure~\ref{fig:all_locs} (Appendix~\ref{app:signal}; the other signal and background parameters are as in Eq.~\eqref{alpha:Higgs}). The relative errors are comparable in each case, indicating that the quality of signal parameter inference does not strongly depend on the signal location.
Overall, the ability to recover the signal position remains good across a range of experimental setups, demonstrating that PoLoN-SB can identify and characterize a localized signal without requiring its position to be approximately known \emph{a priori}.

\subsection{Application to real-world count data}
\begin{figure}[ht!]
    \centering
    \includegraphics[width=\textwidth]{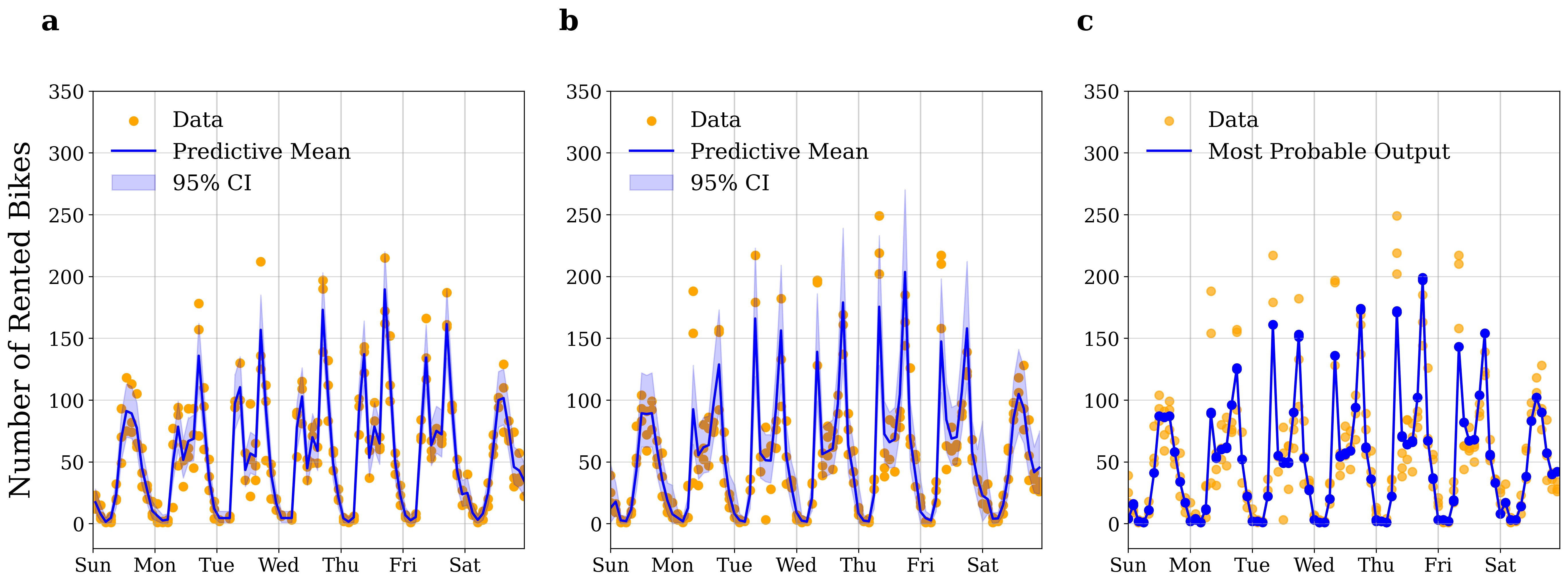}
    \caption{\textbf{PoLoN modeling of the bike rental dataset.} In all three panels, yellow dots represent observations (training set in (a), test set in (b) and (c)) and the horizontal axes show the time of the week, starting from the beginning of Sunday. In panels (a) and (b), the blue line shows the mean of the predictive distribution and the shaded blue region indicates the 95\% CI of the prediction. Panel (a) shows the model fit on the training set (three-week bike renting data, odd hours), with optimized hyperparameters $\sigma_l^\star = 2.194$ and $\gamma^\star = 7.191$.
    Panel (b) shows the mean and the 95\% CI of the predictive distribution for the test set (three-week bike renting data, even hours); the hyperparameters are the same as in (a).
    In panel (c), blue points represent the mode of the predictive distribution for the test set; the connecting blue lines are added to guide the eye.
    }
    \label{fig:bike_rent_train}
\end{figure}

\noindent
\textbf{Bike rental counts.}
We have trained the PoLoN model (Section~\ref{sec:methode_1}) on a standard machine learning dataset containing both hourly and daily counts of rental bikes from 2011 to 2012 for the Capital Bikeshare program in Washington, D.C.~(\url{https://www.kaggle.com/datasets/lakshmi25npathi/bike-sharing-dataset})~\cite{Fanaee2014}. We focus on the hourly counts and use the hour of the day $t_h$ and the day of the week $t_d$ as two entries in the input feature vector (the dataset provides several features such as the hour of the day, the day of the week, whether the day was a working day or a holiday, and the weather conditions (cloudy or sunny)). We combine observations for the odd hours in the first three weeks of bike-sharing data to train the PoLoN model (figure~\ref{fig:bike_rent_train}a), checking the accuracy of our predictions against the even-hour bike rental numbers in the same three weeks (figure~\ref{fig:bike_rent_train}b,c).
We have used the RBF kernel with optimized hyperparameters \(\sigma_l^\star = 2.194\) and \(\gamma^\star = 7.191\).

We observe an excellent agreement between the odd-hour training data and the model, with $R^2 = 0.907$ (the $R^2$ values were computed for training set datapoints vs. the predictive mean $\mathcal{M}(t_h,t_d)$ at each hour; see figure~\ref{fig:bike_rent_train}a).
Furthermore, our model demonstrates a reasonable ability to interpolate, with $R^2 = 0.841$ on the test set which comprises bike rental observations in the even hours over the same three-week period (figure~\ref{fig:bike_rent_train}b; the $R^2$ values were computed as for the training set).
We conclude that our model can use fairly limited data to capture and generalize complex signals often observed in real-world datasets.

\noindent
\textbf{Higgs boson discovery.}
In this section, we focus on extracting the Higgs boson signal from the Quantum Chromodynamics (QCD) background.
The discovery of the Higgs boson at the LHC in July of 2012~\cite{ATLAS2012Higgs,CMS2012Higgs} provided crucial experimental confirmation for the mechanism of mass acquisition by fundamental particles in the Standard Model of particle physics~\cite{Bass2021_HiggsOutlook}.
We have used the open dataset made available by the ATLAS collaboration at LHC and employed the selection criteria as documented in Ref.~\cite{ATL-OREACH-PUB-2020-001}
to create a di-photon invariant mass distribution, $m_{\gamma \gamma}$. The $m_{\gamma \gamma}$ distribution shows the Higgs boson decay as a localized bump superimposed on the smooth background~\cite{ATL-OREACH-PUB-2020-001}.
The Higgs boson dataset provides an ideal test case for the separation of weak localized signals from smoothly varying background distributions.
This task is relevant to many areas of physics and chemistry, with potential applications ranging from exoplanet detection~\cite{Borucki2010Kepler,Ricker2015TESS} to
XPS spectroscopy~\cite{Isaacs2021XPSReview,Krishna2022XPSPedagogy,Bagus2024XPSReview}.

\begin{figure}[!htb]
    \centering
    \includegraphics[width=\textwidth]{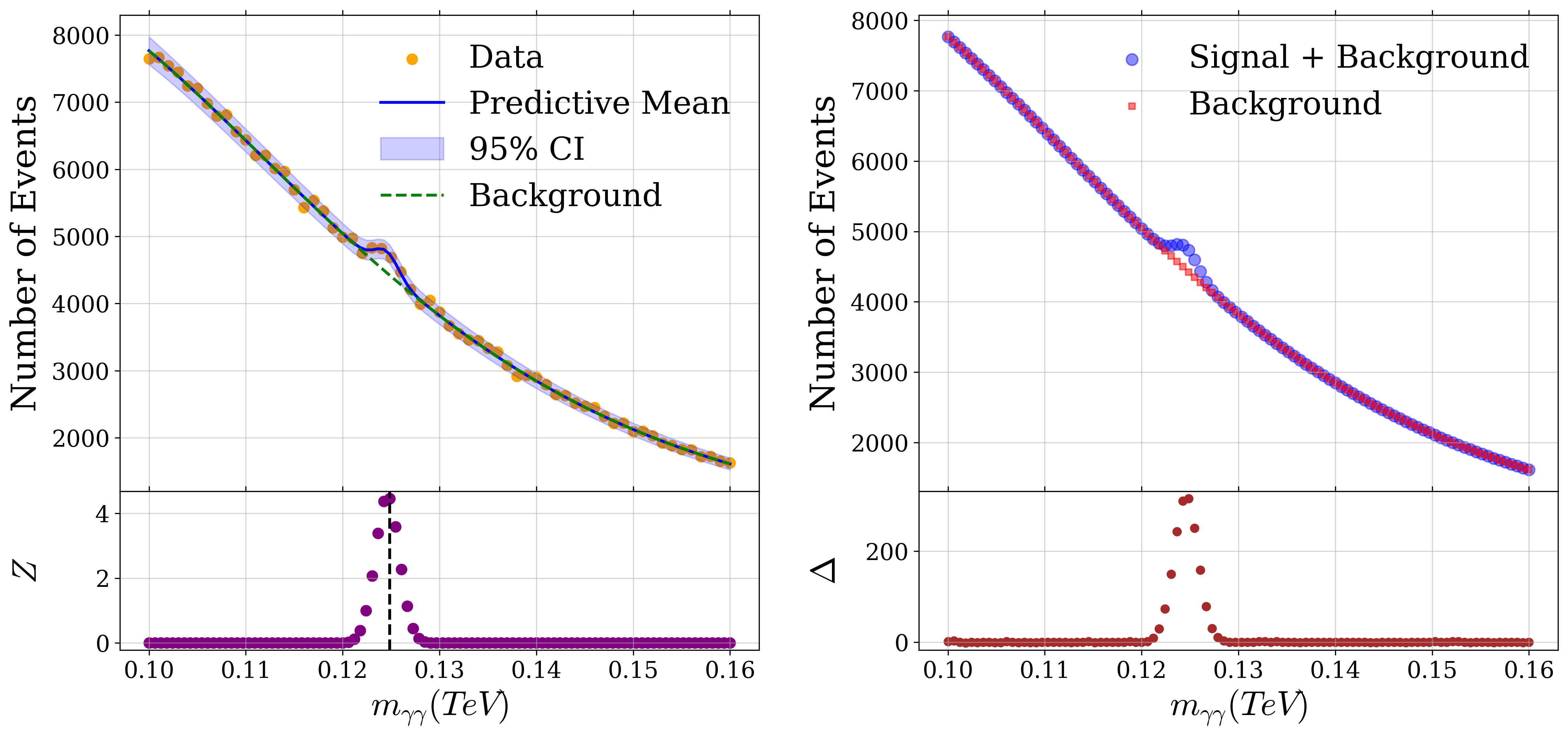}
    \caption{\textbf{Detection of the Higgs boson decay events.} (a) In the upper panel, yellow dots represent the number of events observed in each energy bin.
    In total, there are $M=61$ equal-width (1 GeV) bins in the training data, with bin centers spanning the $[0.10,0.16]\text{ TeV}$ range. The predictive distributions are evaluated on a grid of $100$ equally spaced points over the same range.
    The mean of the background-only predictive distribution (fitted on the data outside of the $[0.12,0.13] \text{ TeV}$ signal window) is shown as a dashed green curve. The mean of the signal+background predictive distribution is shown as a solid blue curve, with the 95\% CI highlighted in light blue. In the lower panel, we plot the Z-scores with respect to the background-only predictive distribution: $Z = V_{b}^{-1/2} (\mathcal{M}_{sb} - \mathcal{M}_{b})$, where $\mathcal{M}_{sb}, \mathcal{M}_{b}, V_{b}$ are the mean of the signal+background predictive distribution and the mean and the variance of the background-only predictive distribution, respectively. The vertical dashed line indicates the location
    of the maximum Z-score: $Z_{max} = 4.45$ at $m_{\gamma \gamma}^\star = 0.1248 \text{ TeV}$.
    (b) In the upper panel, blue circles represent the mode $\widetilde{\mathcal{M}}_{sb}$ of the signal+background predictive distribution; red squares denote the mode $\widetilde{\mathcal{M}}_{b}$ of the background-only predictive distribution. In the lower panel, we plot the difference between the two MAP predictions:
    $\Delta = \widetilde{\mathcal{M}}_{sb} - \widetilde{\mathcal{M}}_{b}$.
    }
    \label{fig:hbdata}
\end{figure}

We employ PoLoN-SB (Section~\ref{sec:methode_2}) since it shows better performance in capturing the overall shape of the signal+background curve (figure~\ref{fig:rmse_comparison}) and is capable of extracting the parameters of the Gaussian signal (figure~\ref{fig:gaussian_parameters}). As in our previous work~\cite{gandrakota2023}, we start by assuming that the window in which the Higgs signal is located is approximately known. Consequently, we exclude the $0.12 \le m_{\gamma \gamma} \le 0.13 \text{ TeV}$ region and fit the background-only model (dashed green line in figure~\ref{fig:hbdata}a, upper panel). This background-only model yields optimized hyperparameters \(\sigma_l^\star = 0.108\) and \(\gamma^\star = 21.253\) for the RBF kernel.

These hyperparameters, which represent the properties of the background, are then used to estimate the parameters of the Gaussian signal, yielding \(S^\star = 321.675\), \(q^\star = 0.12457~\text{TeV}\), and \(u^\star = 0.001247~\text{TeV}\). The mean and the 95\% CI of the full signal+background predictive distribution are displayed in the upper panel of figure~\ref{fig:hbdata}a.
The lower panel shows the Z-scores with respect to the background-only model at each of the 100 evaluation points, which clearly demonstrate the background-signal separation (figure~\ref{fig:hbdata}a, lower panel). The Z-score reaches a maximum of 4.45 at $m_{\gamma \gamma} = 0.1248 \text{ TeV}$, indicating that the signal is highly statistically significant.

The modes (MAP predictions) of the background-only and signal+background predictive distributions are shown in the upper panel of figure~\ref{fig:hbdata}b, with the lower panel showing the difference $\Delta$ between the two modes. We estimate the total number of signal events due to the Higgs boson decay to be $1657$ in the
$[120, 130] \text{ GeV}$
signal window; only one of the $\Delta$ values is negative in this range.
Finally, we have applied the unknown-signal-location procedure of Section~\ref{sec:results_unknown_position} to the Higgs data.
As with the synthetic datasets, we have scanned over a set of sliding
windows of ten-bin width, with a one-bin step (recall that there are
$M=61$ 1 GeV bins in the training data).
The scan over multiple putative signal windows has resulted in
$\bar{S} = 337.280$, $\bar{q} = 0.12456~\text{TeV}$, and $\bar{u} = 0.001392~\text{TeV}$.
These numbers are in excellent agreement with those based on a single
$[0.12,0.13] \text{ TeV}$ signal window,
whose position is informed by the value of $m_{\gamma \gamma}$
expected from the Standard Model.

\section{Discussion and Conclusion}

Non-parametric models based on Gaussian processes provide a flexible framework for describing smoothly varying latent functions~\cite{Rasmussen2005,Bishop2006}. Their combination with Poisson likelihoods represents a natural approach for modeling non-negative integer-count data. In this work, we have modeled the logarithm of the Poisson rate as a Gaussian process, leading, under the Laplace approximation, to the explicit Poisson Log-Normal predictive distribution~\cite{izsak2008maximum,shaban2018poisson}.

We have demonstrated that our computational approach, which we call the Poisson
Log-Normal (PoLoN) process, can be used to make novel predictions and extract the underlying Poisson rate functions from 1D and 2D synthetic datasets.
The predicted Poisson rate functions can be interpreted as de-noised representations of the observed integer count data. We have also used the PoLoN framework to make predictions of the hourly numbers of bike rentals observed in 2011-2012 for the Capital Bikeshare program in Washington, D.C.

In this work, we have focused in particular on the problem of decomposing observed binned event counts into contributions due to a localized signal and a smoothly varying background. As discussed in the Introduction, this problem occurs in many areas of science and often involves low-dimensional datasets in which observations are recorded as a function of 1D or 2D position, time, energy, wavelength, or frequency. To address the problem of signal-background separation, we have developed a version of PoLoN, which we call PoLoN-SB, in which the signal contribution to the Poisson rates is modeled explicitly by a function with a few additional fitting parameters.

The PoLoN-SB approach proceeds in two stages. First, we model the background distribution by excluding the signal window (whose position is assumed to be known for now), obtaining kernel hyperparameters for the background-only model. Second, we keep the kernel hyperparameters fixed and fit the signal parameters on the entire dataset, using an explicit functional form to represent the signal. In this way, we achieve robust decomposition of the total set of observations into background and signal components. After checking the performance of PoLoN-SB on an extensive collection of synthetic datasets with varying signal strengths and dataset sizes, we use our approach to separate Higgs boson decay events from the QCD background on a dataset similar to that used to report the Higgs boson discovery at the LHC in 2012~\cite{ATLAS2012Higgs}. We find that, in line with previous analyses~\cite{ATLAS2012Higgs,CMS2012Higgs,gandrakota2023}, PoLoN-SB assigns high statistical significance to the Higgs boson signal.

In many cases however, the location of the signal window is not known in advance.
To address this challenge, we have extended the PoLoN-SB procedure to a set of sliding
windows, each of which can potentially contain the signal. This has allowed us to predict the signal parameters without assuming the position of the signal-containing window. We have tested this PoLoN-SB extension on both synthetic data and the Higgs boson dataset, obtaining predictions whose quality is comparable to those
based on the signal window known \emph{a priori}.

There are several principal directions in which our computational framework can be  extended in the future.
A natural next step would be to allow searches for multiple, potentially overlapping localized signals rather than a single Gaussian peak.
It would also be of interest to explore replacing the Poisson distribution with the negative binomial distribution, which would allow for modeling datasets with overdispersion~\cite{White1996,CameronTrivedi1998,Hilbe2011}.
While in this work we relied principally on the RBF kernel, future work could explore automated kernel engineering and selection~\cite{pmlr-v28-duvenaud13,pmlr-v28-wilson13}.
For datasets with multi-dimensional feature vectors, automatic relevance determination (ARD) could be used to identify and select the most important features, reducing the effective dimensionality of the data and thus mitigating computational costs~\cite{Rasmussen2005,Bishop2006}. Finally, one could explore the scalability of the PoLoN approach to larger and/or higher-dimensional datasets using sparse Gaussian processes~\cite{QuinoneroCandelaRasmussen2005,Titsias2009,HensmanFusiLawrence2013}.

In conclusion, we have developed the PoLoN predictive framework and used it to separate localized signals from smoothly varying backgrounds in integer-count data. The PoLoN-SB formulation provides an explicit signal-background decomposition, with the sliding-window extension allowing to search for signals whose locations are not known \emph{a priori}. Validation on synthetic and real-world datasets demonstrates the ability of our approach to recover weak localized signals without requiring an explicit parametric model for the background. Together with future developments in multiple-signal searches, likelihood models, kernel selection, feature selection, and scalable GP inference, our framework will provide a foundation for applying GP-based count modeling to a broad range of scientific problems.

\begin{acknowledgments}
A.S. and A.V.M. acknowledge financial and logistical support from the Center for Quantitative Biology, Rutgers University. A.G. is supported by the DOE Office of Science, Award No. DE-SC0023524, FermiForward Discovery Group, LLC under Contract No. 89243024CSC000002 with the U.S. Department of Energy, Office of Science.
The authors are grateful to the Office of Advanced Research Computing (OARC) at Rutgers University for providing access to the Amarel cluster and associated research computing resources. The authors also acknowledge the work of the ATLAS Collaboration to record or simulate, reconstruct, and distribute the Open Data used in this paper, and to develop and support the software with which it was analyzed.
\end{acknowledgments}

\section*{Data Availability Statement}

\noindent
The PoLoN and PoLoN-SB software is written in Python and is available at
\url{https://github.com/AnushkaSaha25/PoLoN-Process}. The GitHub repository also contains the associated scripts and datasets.

\appendix

\renewcommand{\thefigure}{\thesection\arabic{figure}}
\renewcommand{\thetable}{\thesection\arabic{table}}
\renewcommand{\theequation}{\thesection.\arabic{equation}}
\renewcommand{\theHfigure}{\thesection.\arabic{figure}}
\renewcommand{\theHtable}{\thesection.\arabic{table}}
\makeatletter
\@addtoreset{figure}{section}
\@addtoreset{table}{section}
\@addtoreset{equation}{section}
\makeatother

\section{Validation of the PoLoN predictive model on synthetic datasets}
\label{app:validation}

Here, we test the ability of PoLoN to recover smoothly varying
integer-count rates for different functional forms and input
dimensions.

\subsection{1D synthetic datasets}
\label{sec:app_1d}

For each 1D dataset, we select $M=50$ equally spaced values
of the input variable $x$ in the $[1,50]$ range and generate $N_p = 10$ independent
Poisson-distributed observations at each input location. Thus, the resulting
training set contains $N=M N_p=500$ observations. The kernel
hyperparameters are obtained by maximizing the Laplace-approximated
marginal log-likelihood with the L-BFGS-B algorithm, as described in Materials and Methods.

\begin{figure}[!htb]
    \centering
    \includegraphics[width=0.85\textwidth]{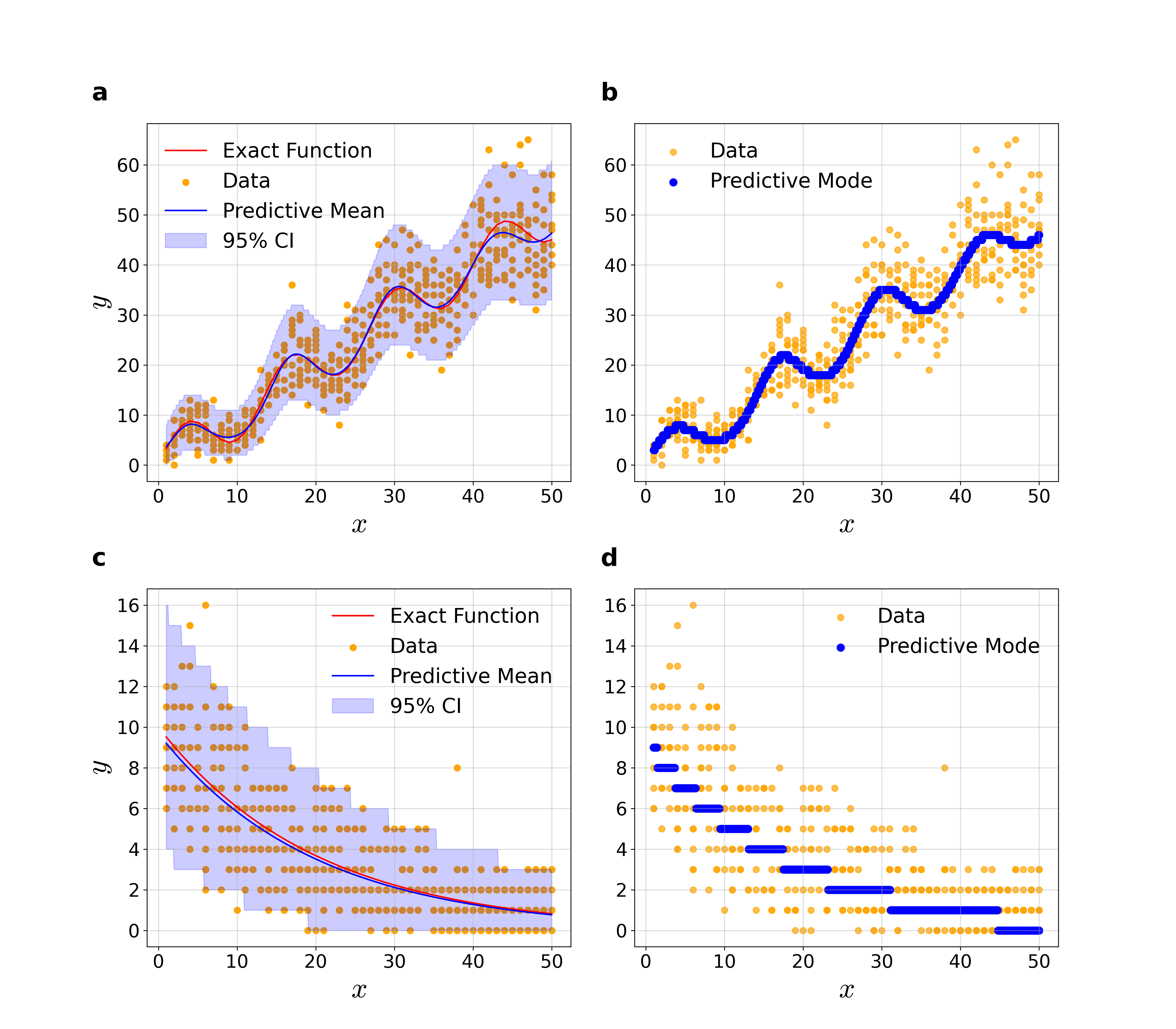}
    \caption{\textbf{PoLoN predictions for 1D integer-count
    data.}
    Panel (a) shows exact Poisson rates $\alpha(x)$ that follow a linear trend with superimposed oscillations (Eq.~\eqref{eq:linsin}; red curve). Training set datapoints are shown as yellow dots. Blue curve: mean of the predictive PLN distribution, light blue shaded area: 95\% confidence interval (CI).
    Panel (b) shows the training set datapoints (yellow dots) and the modes of the predictive PLN distributions for $W=500$ values of $x$ equally spaced in the training dataset range. The modes represent maximum posterior probability (MAP) predictions.
    Panels (c-d) -- same as (a-b), but for the exponentially decaying $\alpha(x)$ in Eq.~\eqref{eq:exp}.
    }
    \label{fig:lin_sin_2}
\end{figure}

We first consider a Poisson rate consisting of a linear trend with
superimposed oscillations,
\begin{equation}
    \alpha(x) = ax + A\sin(2\pi\nu x),
    \label{eq:linsin}
\end{equation}
with $a=1$, $A=5$, and $\nu=0.075$.
Figure~\ref{fig:lin_sin_2}a shows the training data together with the
exact Poisson rate and the mean of the predictive PLN
distribution; the 95\% confidence interval
quantifies the predictive uncertainty.
Figure~\ref{fig:lin_sin_2}b
shows the corresponding modes of the predictive distribution, which
represent the most probable integer-count predictions.
We employ an RBF kernel with optimized hyperparameters
$\sigma_l^\star=7.170$ and $\gamma^\star= 8.452$ (cf. Eq.~\eqref{RBF:kernel}).
We find an excellent overall agreement between predicted and exact Poisson rates, with the root-mean-square error (RMSE) $\delta = 0.862$ over $W=500$ equally spaced points in the $[1,50]$ training data range (figure~\ref{fig:lin_sin_2}a), where $\delta = W^{-1/2}\sqrt{\sum_{j=1}^W [\alpha(x_j) - \mathcal{M}(x_j)]^2}$
($\mathcal{M}(x_j)$ is the mean of the predictive distribution at position $x_j$
and $\alpha(x_j)$ is the exact Poisson rate).

\begin{figure}[!htb]
    \centering
    \includegraphics[width=0.85\textwidth]{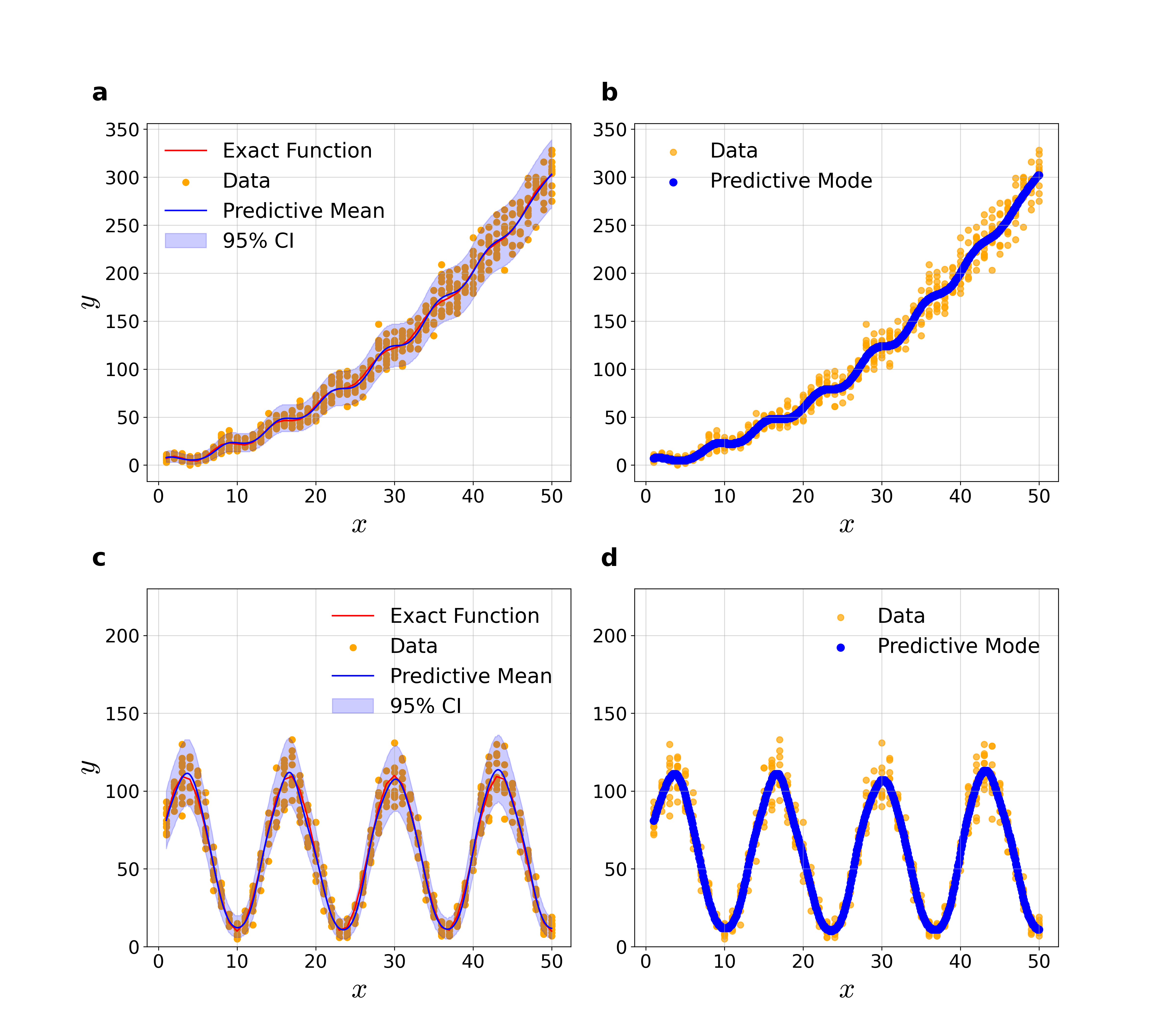}
    \caption{\textbf{Additional PoLoN predictions for 1D
    integer-count data.}
    Panels (a–b): same as figure~\ref{fig:lin_sin_2}a,b, but for
    $\alpha(x)$ following a quadratic trend with superimposed oscillations (Eq.~\eqref{eq:quadsin}; red curve).
    Panels (c-d): same as figure~\ref{fig:lin_sin_2}a,b, but for
    $\alpha(x)$ following a purely oscillatory trend (Eq.~\eqref{eq:sin}; red curve).
    }
    \label{fig:quad_sin}
\end{figure}

Next, we consider an exponentially decaying Poisson rate,
\begin{equation}
    \alpha(x)=A e^{-\kappa x},
    \label{eq:exp}
\end{equation}
with $A=10$ and $\kappa=0.05$.
Because the logarithm of
this rate is linear in $x$, we use a linear kernel, Eq.~\eqref{lin:kernel};
the optimized parameters are $\theta_1^\star=0.0026$ and $\theta_2^\star=7.195$.
As shown in figure~\ref{fig:lin_sin_2}c,d, the predictive mean is in
close agreement with the exact Poisson rate, with an RMSE of
$\delta=0.169$ over $W=500$ equally spaced evaluation points. This
example is particularly relevant to the signal--background studies
in the main text because an exponentially varying component is used
as the smooth background in the synthetic signal datasets.

Third, we consider a Poisson rate containing both a quadratic trend and
sinusoidal modulation,
\begin{equation}
    \alpha(x) =
    ax^2 + bx + c + A\sin(2\pi\nu x),
    \label{eq:quadsin}
\end{equation}
with $a=0.1$, $b=1$, $c=2$, $A=5$, and $\nu=0.15$.
We use the RBF kernel with optimized hyperparameters
$\sigma_l^\star=4.113$ and $\gamma^\star=10.128$.
As shown in figure~\ref{fig:quad_sin}a,b, the PoLoN predictive distribution
captures both the broad quadratic variation and the superimposed
oscillatory structure. The RMSE of the predictive mean is
$\delta=1.803$ over $W=500$ equally spaced evaluation points.

Finally, we test the PoLoN model using a purely oscillatory Poisson rate,
\begin{equation}
    \alpha(x) = A\sin(2\pi\nu x)+k,
    \label{eq:sin}
\end{equation}
with $A=50$, $\nu=0.075$, and $k=60$.
As before, the RBF kernel was used, with
optimized hyperparameters $\sigma_l^\star=2.869$ and $\gamma^\star= 9.958$.
Figure~\ref{fig:quad_sin}c,d shows that the predictive distribution
reproduces the periodic variation of the underlying Poisson rate.
The resulting RMSE is $\delta=2.408$ over $W=500$ equally spaced
evaluation points.

Together, these 1D examples demonstrate that PoLoN can
recover Poisson rate functions of substantially different functional
forms, including polynomial, oscillatory, and exponentially varying
rates.

\begin{figure}[!htb]
    \centering
    \includegraphics[width=\textwidth]{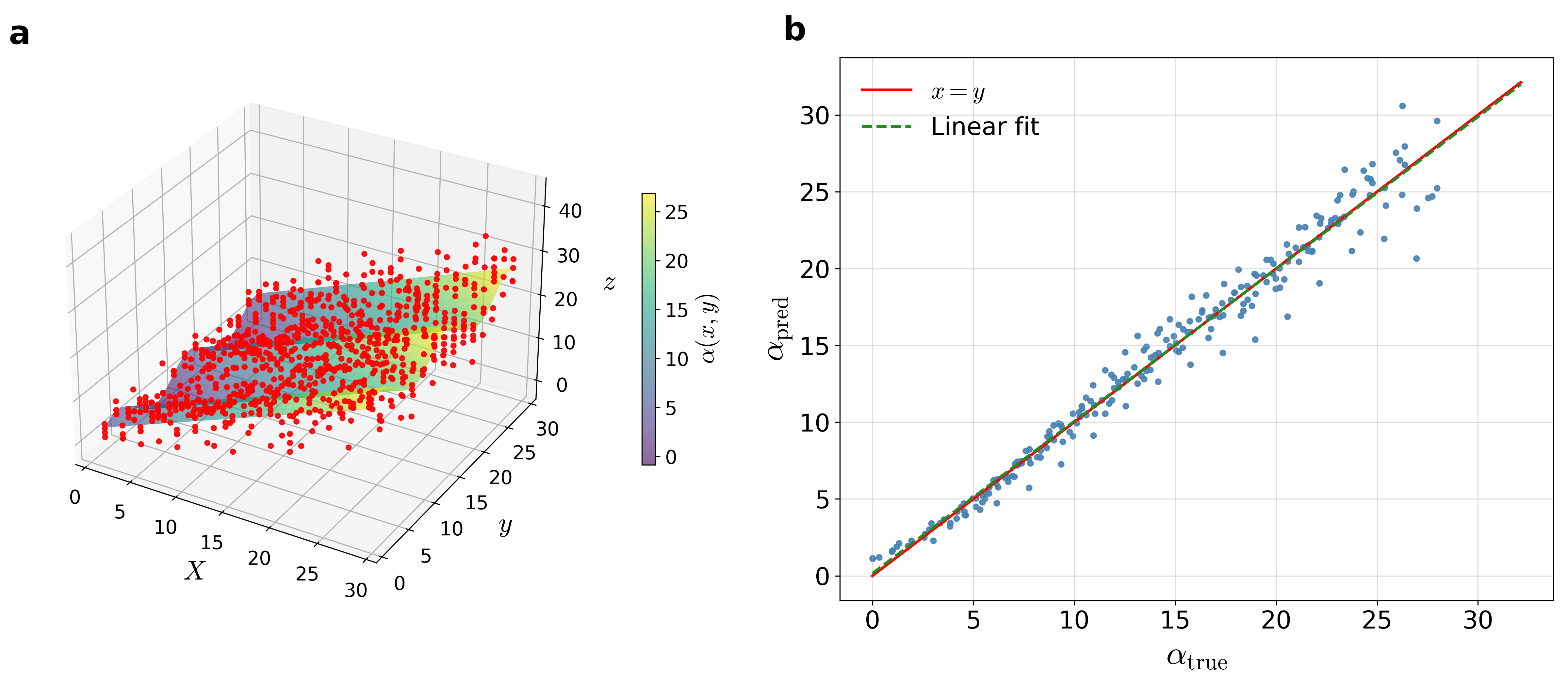}
    \caption{\textbf{PoLoN predictions for 2D integer-count
    data.}
    Panel (a) shows the 2D Poisson rate function $\alpha(x, y)$
    with a linear trend in the $x$-direction and oscillations in the $y$-direction (Eq.~\eqref{eq:twod}), for $x,y \in [1, 30]$. The training datapoints are shown in red.
    Panel (b) shows a comparison between predicted and actual Poisson rates for $225$ $(x,y)$ pairs in the test set.
    The solid red line is the ``$x=y$'' line; the dashed green line is the least-squares fit.
    The coefficient of determination is $R^2 = 0.972$.
    }
    \label{fig:2d}
\end{figure}

\subsection{2D synthetic dataset}
\label{sec:app_2d}

Here, we consider a Poisson rate of the form
\begin{equation}
    \alpha(x,y)=ax+b\sin(cy)+d,
    \label{eq:twod}
\end{equation}
with $a=0.8$, $b=3.0$, $c=0.5$, and $d=1.0$. This function contains a
linear dependence in the $x$ direction and an oscillatory dependence
in the $y$ direction.

The training inputs consist of a $15\times15$ grid ($M=225$) with
$x,y\in[1,29]$, corresponding to the values
$x_i,y_j=\{1,3,\ldots,29\}; i,j=1 \ldots 15$. Four independent Poisson observations
are generated at each grid location ($N_p=4$), resulting in
$N=900$ training observations.
For testing, we have constructed a separate $15\times15$ grid with
$x,y\in[2,30]$ (i.e., $x_i,y_j=\{2,4,\ldots,30\}; i,j=1 \ldots 15$). Thus, the test locations are interspersed with the
training locations but were not used during training. We use an RBF
kernel with the optimized hyperparameters
$\sigma_l^\star=6.572$ and $\gamma^\star=6.406$.
Figure~\ref{fig:2d}a shows the exact two-dimensional Poisson rate
together with the training datapoints. Figure~\ref{fig:2d}b compares
the predicted and exact Poisson rates at the 225 test locations. The
predictions show excellent agreement with the underlying rate, with
$R^2=0.972$.

\section{Signal-background separation in synthetic datasets}
\label{app:signal}

Here, we provide additional figures for the signal-background separation analysis in Sections~\ref{sec:results_signal_background} and \ref{sec:results_unknown_position}.

\begin{figure}[!htb]
    \centering
    \includegraphics[width=0.65\textwidth]{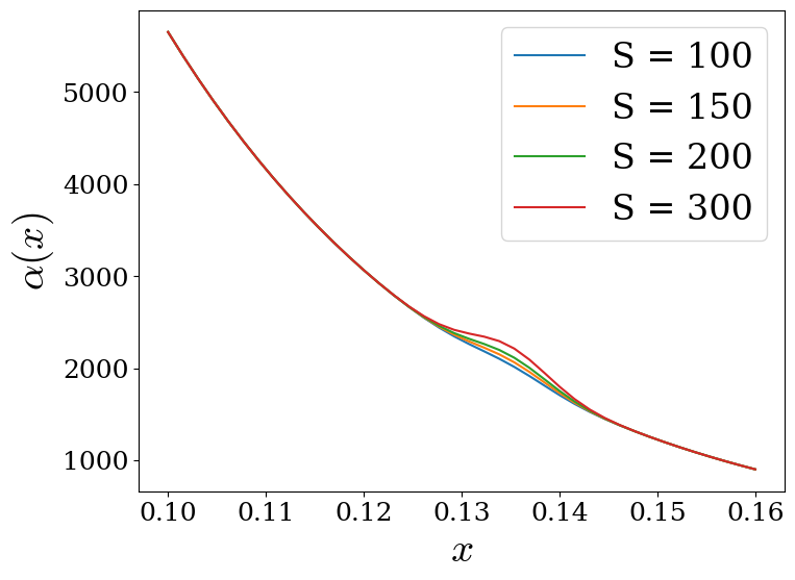}
    \caption{\textbf{Poisson rate functions for separating signal from background in synthetic datasets.} Representative examples of Poisson rate functions $\alpha(x)$, Eq.~\eqref{alpha:Higgs}, for the exponentially decaying background with a Gaussian signal of varying strength $S$.
    }
    \label{fig:alpha_signal}
\end{figure}

\begin{figure}[!htb]
\centering
\includegraphics[width=0.85\textwidth]{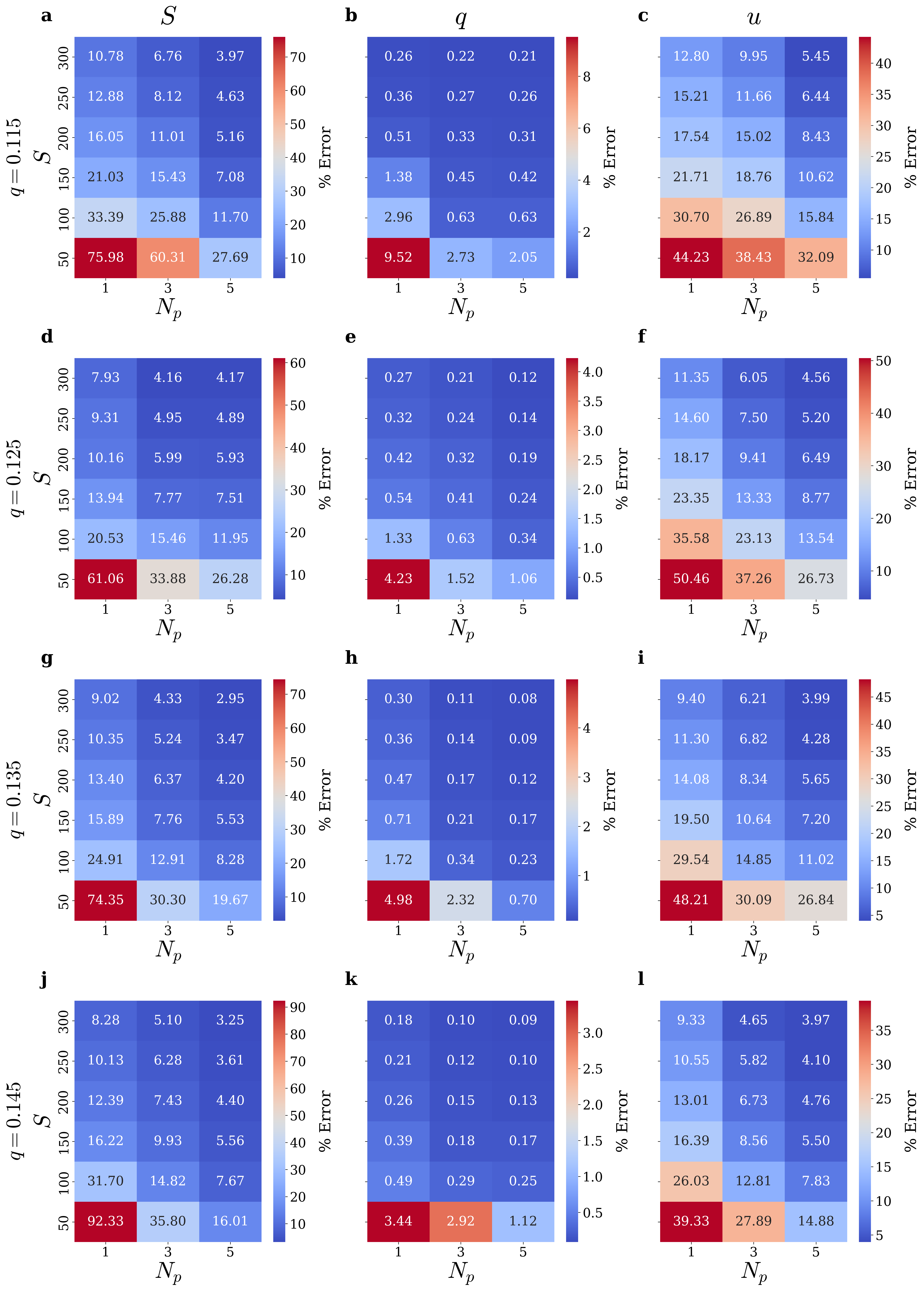}
\caption{\textbf{Relative percentage errors in Gaussian signal parameter recovery: multiple signal locations, unknown signal position.}
Same as figure~\ref{fig:unknown_position_q0135}, for $q = (0.115,~0.125,~0.135,~0.145)$
(the $q = 0.135$ results from figure~\ref{fig:unknown_position_q0135} are reproduced here for completeness).
}
\label{fig:all_locs}
\end{figure}

\clearpage
\bibliography{biblography}

\end{document}